\documentclass[a4paper,12pt]{article}
\usepackage{setspace}
\usepackage{codams,float,amssymb}
\usepackage[leqno]{amsmath}
\usepackage[dvipdf]{graphicx}
\floatstyle{ruled}
\newfloat{displaycode}{htp}{alg}[section]
\floatname{displaycode}{Eiffel code}

\newtheorem{definition}[theorem]{Definition}
\newtheorem{lemma}[theorem]{Lemma}

\newtheorem{proposition}[theorem]{Proposition}
\newtheorem{corollary}[theorem]{Corollary}

\numberwithin{equation}{section}

\newcommand{\be}{\begin{equation*}}
\newcommand{\ee}{\end{equation*}}
\newcommand{\bg}{\begin{gather*}}
\newcommand{\eg}{\end{gather*}}
\newcommand{\hb}{\hfil\break}

\DeclareGraphicsExtensions{.eps}
\title{
Compact families of Jordan curves and
convex hulls in three dimensions\footnote{These results were presented
at EuroCG 2015, Ljubljana, Slovenia, in March 2015.}
}
\author{Colm \'O D\'unlaing\\
Mathematics, Trinity College, Dublin 2, Ireland
\thanks{e-mail: odunlain@maths.tcd.ie.
Mathematics department website http://www.maths.tcd.ie.}}

\begin{document}

\maketitle

\begin{abstract}
We prove that for certain families of semi-algebraic convex bodies
in $\IR^3$,
the convex hull of $n$ disjoint bodies
has $O(n\lambda_s(n))$ features, where $s$ is a constant depending
on the family:
$\lambda_s(n)$ is
the maximum length of order-$s$ Davenport-Schinzel sequences
with $n$ letters.
The argument is based on an apparently new idea
of `compact families' of convex bodies
or discs, and `crossing content' among discs.
\end{abstract}

\section{Introduction}
\label{sect: introduction}
\numpara
The construction of convex hulls is a well-studied
problem, certainly for finite sets of points
in any dimension, and for more general sets,
such as curved objects in two dimensions [\ref{acv}],
quadric surfaces in three dimensions [\ref{wolpert}],
and spheres in any dimension [\ref{bcddy}].
This paper gives a reasonably straightforward derivation
of an $o(n^2 \log^* n)$ upper bound for the feature
complexity (descriptive complexity) of the convex hull
of $n$ disjoint bodies in three dimensions, granted that
the bodies come from a `compact family,' a term defined
in this paper.

\numpara
In 1995 Hung and Ierardi [\ref{hi}] reported
$O(n^{2+\varepsilon})$\footnote{This is how the
complexity was stated, though probably
an estimate close to ours could have been
given.} complexity bounds, together
with algorithms for constructing the hull,
but their approach is indirect and hard to understand.
In this paper we (hopefully) develop a theory sufficient
for a convincing proof.

\begin{figure}
\centerline{\includegraphics[height=1.5in]{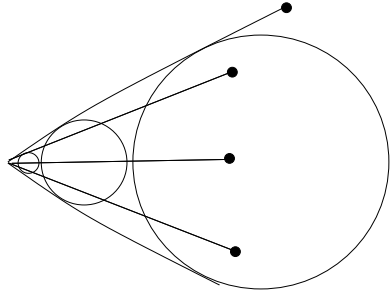}}
\caption{$n$ spheres, $\Omega(n^2)$ features.}
\label{fig: michelin}
\end{figure}

\numpara
$S$ will be a set of $n$ disjoint convex bodies in
$\IR^3$.
\be
H(S)
\ee
denotes the convex hull of $S$.  As in [\ref{wolpert}]
the boundary $\partial H(S)$ is divided into
{\em exposed facets}, {\em tunnel facets}, and {\em planar facets}.
These, with their separating edges and vertices, constitute
the features of $H(S)$. In the case of spherical bodies
it is known that $H(S)$ has $O(n^2)$ features, and this
is also a lower bound
(Figure \ref{fig: michelin}, [\ref{ss},\ref{bcddy}]).\footnote{
This construction is possible with spherical bodies of radius $r$,
where $1 \leq r \leq 2$, say, i.e., the lower bound
holds for compact
families such as are discussed in this paper.}

Every facet is incident to an edge or vertex
of an exposed facet, so
the feature complexity can be estimated by 
counting the edges and/or vertices on the exposed
facets.  Thus the complexity can be reduced to that
of unions of discs.

\begin{figure}
\centerline{\includegraphics[height=1in]{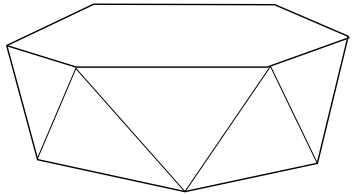}}
\caption{The convex hull of two polygons can have high complexity.}
\label{fig: drum}
\end{figure} 

\numpara
It is necessary to assume some complexity bounds on the
bodies.  For example, Figure \ref{fig: drum} shows how
the convex hull of two bodies can have many features.
To eliminate this we assume that the bodies are semialgebraic
of bounded degree.

Unions of $n$ circular discs have complexity $O(n)$,
whereas unions of $n$ thin ellipses can
have complexity $\Omega(n^2)$, obviously because they
are `thin,' and the analysis of various notions of
`fatness' which reduce the complexity, has been of great interest
[\ref{dck},\ref{ek}].

\begin{figure}
\centerline{\includegraphics[height=1in]{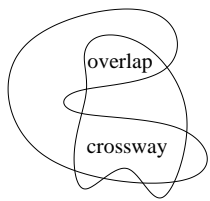}}
\caption{An overlap has two sides/vertices; a crossway has more.}
\label{fig: overlap}
\end{figure}

One distinguishes two kinds of disc intersection:
overlaps and crossways. Given two (topological) discs
$D_1$ and $D_2$, an {\em overlap} (respectively,
{\em crossway}) is a connected component of
$D_1 \cap D_2$ whose intersection with the boundaries
$\partial D_1$ and $\partial D_2$ is connected
(respectively, disconnected): see Figure \ref{fig: overlap}.

Given a list of $n$ discs where any two intersect
in at most one component, and that an overlap,
the arrangement is termed one of {\em pseudodiscs}
and the union has $O(n)$ features [\ref{dck}].

On the other hand, $n$ thin ellipses can have
$\Omega(n^2)$ crossways.

In order to limit the number of crossways, we
develop idea of {\em positive crossing content,}
where there is a positive lower bound on the area of any crossway.

We show that, given positive crossing content and
bounded intersection (a bound
on the number of intersection components between any
two discs), the union has $O(n)$ overlaps,
by planarity arguments; crossways are handled differently.

\begin{figure}
\centerline{\includegraphics[height=1in]{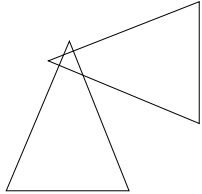}}
\caption{kinks will rule out positive crossing content.}
\label{fig: kinks}
\end{figure}

Positive crossing content
requires the disc boundaries to be {\em differentiable}
(Figure \ref{fig: kinks}). The `compact families' of discs
studied in this paper have continuously differentiable boundaries,
and we prove, using compactness and continuity arguments,
that such families have positive crossing content.
This is our version of `fatness': possibly `stiffness' is
a better word, indicating that the disc boundaries are resistant
to kinks (Figure \ref{fig: kinks}).

\numpara
Accordingly, our point of departure is the notion of
a {\em compact family of convex bodies}, which have
twice-differentiable boundaries and have
a distance function based on the $C^2$ norm. From these we pass
to compact families of discs which are $C^1$ and
have a metric based on the $C^1$ norm. We show that
the map from bodies to discs --- which are hidden
regions on the bodies' boundaries --- is continuous, from which
the compactness of the disc family and positive crossing
content are derived.


From positive crossing content we can show that
on any body $B$ there are $O(n)$ pairs $(D,E)$ of
incident hidden and exposed areas (which we call discs and holes),
whence the exposed areas on $B$
have $O(\lambda_s(n))$ features, and $H(S)$ has
$O(n \lambda_s(n))$ features overall. Here $\lambda_s(n)$ is
the maximum length of $n$-letter order-$s$ Davenport-Schinzel sequences,
and $s$ is a constant
depending on the semialgebraic complexity of
the bodies.  There are asymptotically exact formulae [\ref{as}] for
$\lambda_s(n)$, which are slightly convoluted;
$O(n^2 \log^* n)$ is a relatively simple
over-estimate for the feature complexity of $H(S)$.

\subsection{Concluding remarks and a possible development}

The combinatorial side of this paper is fairly
straightforward while achieving good results based on reasonable
assumptions of differentiability and so forth.  This is probably
one of the first studies of differentiable functions with 
combinatorial complexity in view. For this reason one must
be careful with the continuum mathematics.  There are two
important results here.  First, the property of positive crossing
content for compact families of Jordan curves. Second, that
pre-seams form such a compact family.  Our proof of these
two facts is long, especially the latter, but that seems to be necessary
to put the theory on a solid footing.

A consequence of bounded crossing content is that there must
be $O(1)$ pairwise disjoint crossways.  This is a severe
restriction, but not so severe as to prevent us deducing the feature
complexity of convex hulls.
But the theory should be applicable to
non-compact families of discs in the plane
with some relativised form of bounded crossing content.

\section{Metric spaces; differentiation}
\label{sect: metric}
\subsection{Metrics}
A metric space is a set $X$ together with a distance
function $d: X \times X \to [0,\infty)$ such that
$d(x,y) = d(y,x)$, $x = y \iff d(x,y)=0$, and
$d(x,z) \leq d(x,y)+d(y,z)$.  This gives rise in the
usual way to a topology on $X$.

A topological space is compact if every open covering contains a finite
subcover.  For metric spaces, compactness and sequential compactness
are equivalent; a metric space $X$ is sequentially compact if every
infinite sequence $x_n \in X$ contains a convergent subsequence.

Euclidean distance gives a metric on $\IR^n$:
\be
d(x,y) = \| x-y \| .
\ee

If $X$ is a nonempty compact subspace of $\IR^n$ and $Y$ is a set
of continuous functions on $X$, then
\be
d(f,g) = \sup_{x \in X} \| f(x) - g(x) \|
\ee
is a metric on $Y$.
We have various alternative notations for this `sup norm.' One
is
\be
\| f-g \|_\infty ,
\ee
another is
\be
\| f-g \|_X ,
\ee
and another (for the `$C^2$ norm')
\be
\| f-g \|_{C^2}.
\ee

\subsection{Neighbourhoods}
\label{subsec: neighbourhoods}

We use a uniform notation for neighbourhoods with respect
to different metrics.  Suppose that $X$ is a metric space
with metric $d$.

Given $r > 0$ and $x \in X$, the {\em open $r$-neighbourhood}
of $x$ is
\be
N_r (x) = \{ y \in X:~ d(x,y) < r \}
\ee
and the {\em closed $r$-neighbourhood} of $x$ is
\be
\overline{N_r} (x) = \{ y \in X:~ d(x,y) \leq r \} .
\ee
Although the definition of neighbourhood involves the
metric $d$, the notation does not mention $d$ explicitly.

\subsection{Differentiability}
\label{subsec: differentiability}

Let $V$ be an open subset of $\IR^n$ and $f: V \to \IR^m$
a function. Given $x \in V$, $f$ is {\em differentiable
at $x$} if there exists a matrix $A_{m\times n}$ such that
\be
f(x+h) = f(x) + Ah + o(\| h\|)
\ee

\numpara
\label{par: frechet derivative}
In this case, $A$ is unique, and it is called the
{\em Fr\'{e}chet derivative} of $f$ at $x$, abbreviated
$f'(x)$.  The map $f$ is {\em continuously differentiable}
on $V$ if it is differentiable everywhere in $V$ and
the map $x \mapsto f'(x)$ is continuous [\ref{spivak}].
In this case, the derivative itself may be continuously
differentiable and $f$ is twice differentiable, and we
write $f''(x)$ for the second derivative.

When $m=1$, i.e., the maps are into $\IR$,
we write $C^1(V)$, $C^2(V)$ for the family of continuously
differentiable or twice-differentiable maps from $V$ to $\IR$.
With $n=3$ and $m=1$,
\begin{gather*}
f'(\vec{x}) = 
\left [ \begin{array}{ccc}
\frac{\partial f}{\partial x} &
\frac{\partial f}{\partial y} &
\frac{\partial f}{\partial z} \end{array}\right ]
\end{gather*}
The {\em gradient} of $f$ is the transpose of $f'(x)$:
\be
\nabla_f (\vec{x}) =
\left [ \begin{array}{c}
\frac{\partial f}{\partial x} \\[3pt]
\frac{\partial f}{\partial y} \\[3pt]
\frac{\partial f}{\partial z} \end{array}\right ]
\ee
and we identify $f''(\vec{x})$ with a matrix,
the derivative of $\nabla_f(\vec{x})$:
\be
f''(\vec{x}) = 
\left [ \begin{array}{ccc}
\frac{\partial^2 f}{\partial x^2} &
\frac{\partial^2 f}{\partial y \partial x} &
\frac{\partial^2 f}{\partial z \partial x} \\[3pt]
\frac{\partial^2 f}{\partial x \partial y} &
\frac{\partial^2 f}{\partial y^2} &
\frac{\partial^2 f}{\partial z \partial y} \\[3pt]
\frac{\partial^2 f}{\partial x \partial z} &
\frac{\partial^2 f}{\partial y \partial z} &
\frac{\partial^2 f}{\partial z^2}
\end{array}\right ] .
\ee

\section{Convex hulls, their features and their
hidden and exposed regions}
A subset $C$ of $\IR^n$ is {\em convex} if for any
$x,y\in C$ the closed line-segment $xy$ is contained
in $C$, i.e., for all $0 \leq t \leq 1$,
$(1-t)x + t y \in C$.

If $X \subset \IR^n$ then its closed convex hull
\be
H(X)
\ee
is the intersection of all closed convex sets containing $X$.
It is the smallest closed convex set containing $X$.

\numpara
\label{par: assumptions}
Let $S$  be a set of subsets of $\IR^3$.
In this paper they will be referred to as convex
bodies.

We make the following initial assumptions,
which have been invoked in a previous
work [\ref{hoy}].  Further conditions
will be stated in the next section.

\begin{itemize}
\item
The bodies are nonempty, closed, bounded, and convex.
\item
They are in general position:
no four bodies possess a common tangent plane.
\item
They are pairwise disjoint.
\item
They are {\em rounded}\, meaning that
their boundary surfaces have unique tangent planes (or
outward unit normals)
at all points, and every tangent plane meets the
boundary at just one point.
\end{itemize}

\noindent
$H(S)$ is the (closed) convex hull of \, $\bigcup S,$
i.e., of \, $\bigcup\{B:~~B\in S \}$.\hfil\break

{\bf Structure of $H(S)$.}
The {\em features} of $H(S)$ are its facets, edges,
and vertices, as follows.
As discussed in [\ref{wolpert},\ref{hoy}],
$\partial H(S)$ is naturally divided into
connected regions: its
{\em exposed facets},
{\em tunnel facets}, and {\em planar facets}.
The exposed facets are (path-)\,connected components of
$(\partial H(S)) \cap \bigcup S$,
tunnel facets are connected part-surfaces generated by line-segments
touching two bodies, and (since the bodies are in general position)
planar facets are triangular.
Tunnel facets are bounded by two exposed facets and by
two planar facets (or are quasi-cylindrical, joining
two bodies).

\noindent
Facets meet along {\em edges}, and edges meet at {\em vertices};
an edge could be  a closed loop.

\begin{figure}
\centerline{\includegraphics[height=1.5in]{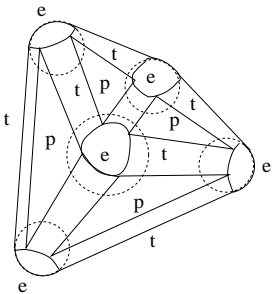}}
\caption{Convex hull of five spheres.
Exposed facets, tunnel facets, and planar facets
are marked e, t, and p, respectively.}
\label{facets.fig}
\end{figure}

Under  the assumption of
general position, no facet touches more than
three bodies.
Figure \ref{facets.fig}
illustrates these features, except that exposed facets need
not be simply connected.

The {\em feature complexity} of $H(S)$ is the total number of
features, generally proportional to the number of facets.

If $B \in S$, we call
\be
\partial B \cap \partial H(S)
\ee
the {\em exposed part} of $B$, whereas
\be
\overline{\partial B \cap H(S)^\circ}
\ee
\noindent is its {\em hidden part}.
(The exposed and hidden parts, according to this
definition, are both closed
and they intersect along their common boundaries).\footnote{
$X^o$ is the interior of $X$,
$\overline{X}$ its closure, and $\partial X =
\overline{X} \backslash X^\circ$ its boundary.}

\section{Compact families of convex bodies and discs, seams
and pre-seams}
\label{sect: compact fam}

\subsection{Compact families of convex bodies}
\label{subsect: compact families of convex bodies}

In addition to the requirements stated
in paragraph \ref{par: assumptions},
our analysis requires further assumptions about the
kinds of body occurring in $S$.  We require that
each is a translated copy of a `model' body.
The `model' bodies are to be taken from a restricted
family.
For this reason, a {\em model} is a convex body
subject to various restrictions.

By the {\em derivative} $f'(x)$ of a function $f$ at $x$ we mean
the Fr\'echet derivative mentioned in (\ref{par: frechet derivative}).

\noindent
A $C^r$-function is one which is $r$ times continuously
differentiable.

We assume that each body in $S$ is specified by
an inequality
\begin{gather*}
f( x-a ) \leq 1:\\
B^{f,a} = \{ x \in \IR^3:~ f(x-a) \leq 1 \}.
\end{gather*}
\noindent $B^{f,a}$ is
the translation by $a$, or a {\em placement}, of a {\em model}
\be
B^f = B^{f,O} =  \{ x: ~ f(x) \leq 1 \}.
\ee
$\cal G$ is the family of all such functions $f$.

Recall our notation for open and closed balls in $\IR^3$
(\ref{subsec: neighbourhoods}):
\begin{gather*}
N_d(x) = \{ y \in \IR^3: ~ \|y-x\| < d \}\\
{\overline{N}}_d(x) = \{ y \in \IR^3: ~ \|y-x\| \leq d \}
\end{gather*}

\numpara
\label{par: compact model assumptions}
In addition to the
assumptions  \ref{par: assumptions},
for every $f\in \cal G$,

\begin{itemize}
\item
$f \in C^2(\IR^3)$:
$f$ is defined and twice continuously differentiable everywhere
in $\IR^3$ (\ref{subsec: differentiability}).
\item
For all $x$ outside $N_{1.5}(O)$, $f(x)$ has the constant
value $2$.
It follows that $B^{f,O}$ is contained in the
open ball $N_{1.5}(O)$.
\item
$f''(x)$ is positive definite, and $f'(x)$ is nonzero, for
all $x$ in $\partial B^{f,O}$, hence for all $x$ in an
open neighbhourhood of $\partial B^{f,O}$.

\item
The origin is interior to all models,
i.e., $f(O) < 1$ for all $f\in \cal G$.

\item
$f$ is piecewise algebraic of bounded degree.
More precisely, for each $f \in \cal G$,
there is a covering $S_1 \cup \ldots \cup S_k$ of $\IR^3$
by semi-algebraic sets, and for $1 \leq i \leq k$ there
is a polynomial $p_i (x,y,z)$, such that
\be
f_{\left | S_i \right .} =
{p_i}_{\left | S_i \right .}.
\ee

\end{itemize}

The norm $\| x \|$ is the usual Euclidean norm, which may
also be used for matrices, and thus for second derivatives.

For each $f \in \cal G$, $f \equiv 2$ outside a compact
set $\overline{N_{1.5}}(O)$, so it is bounded, and we may define its
`sup norm'
\be
\| f \|_\infty =
\sup \{ \|f(x) \|: ~ x \in \IR^3\}.
\ee
The first and second derivatives vanish outside
$\overline{N_{1.5}}(O)$,
so their sup norm is also well-defined,
and we can define

\begin{definition}
\label{def: C2 norm}
The {\em $C^2$ norm} on parametrisations $f\in \cal G$ is
\begin{gather*}
\| f \|_{C^2} =\text{\rm (def)}\quad
\max \left (
\| f\|_\infty, ~
\| \nabla_f \|_\infty, ~
\| f'' \|_\infty \right ).
\end{gather*}
and the $C^2$-distance $d(f,g)$ between two functions
is $\|f-g\|_{C^2}$.
\end{definition}

\begin{definition}
A family of models is {\em compact} if the parametrising
family $\cal G$ is compact under the $C^2$ metric.
\end{definition}

\begin{definition}
\label{def: outward unit normal}
Given a body $B=B^{f,a}$ and $p\in \partial B$,
the (outward) {\em unit normal} $n(p)$ at $p$ is
\be
n_{f,a}(p)\quad\text{\rm or}\quad n(p) = \frac{\nabla_f(p-a)}{\|\nabla_f(p-a)\|}.
\ee
\end{definition}
The subscripts in $n_{f,a}$ will be omitted if no ambiguity arises.

\begin{proposition}
\label{prop: normal homeo}
If $B$ is a rounded compact convex body, then the map
\be
\partial B \to S^2:\quad p \mapsto n(p)
\ee
is a homeomorphism
{\rm [\ref{hoy}, Lemma 1].}\qed
\end{proposition}

\subsection{Compact families of discs}
\label{subsect: compact families of discs}

We shall prove that hidden regions arising from a compact
family of models form a compact family of discs (as defined below).
A transformation will be applied to hidden regions so they
are topological discs on the unit sphere $S^2$.

Suppose $\phi:~ [0,2\pi]\to \IR^3$ is a continuous
map.  By its {\em derivative} $\frac{df}{d\phi}$ at $\phi$ is meant a one-
or two-sided limit, presuming it exists:
\begin{gather*}
\frac{df}{d\phi}=
\begin{cases}
\lim_{h \to 0} \frac{f(\phi+h)-f(\phi)}{h}
\quad\text{if}~ 0 < \phi < 2\pi,\\
\lim_{h \to 0^+} \frac{f(h)-f(0)}{h}\quad\text{if}~ \phi = 0,\\
\lim_{h \to 0^-}
\frac{f(2\pi+h)-f(2\pi)}{h} \quad\text{if}~ \phi = 2\pi.
\end{cases}
\end{gather*}

\begin{definition}
\label{def: C^1 Jordan curve}
A {\em (closed) disc} is generally taken in the topological sense, i.e.,
a topological space homeomorphic to the closed unit disc
\be
\{ (x,y)\in \IR^2: ~ x^2+y^2 \leq 1\} .
\ee
This paper is concerned with discs on the unit sphere $S^2$.
An {\em oriented $C^1$ Jordan curve} in $S^2$ is the image of
a map $f: [0,2\pi] \to S^2$, satisfying the following conditions.
\begin{itemize}
\item
The map $f$ is injective, except that $f(0) = f(2\pi)$.
\item
It is continuously differentiable, i.e., $\frac{df}{d\phi}$ is
defined and continuous everywhere
and $\frac{df}{d\phi} (0) = \frac{df}{d\phi} (2\pi)$.
\item
Its derivative, a vector
in $\IR^3$, is nowhere zero: $\frac{df}{d\phi}\not= \vec{O}$.
\end{itemize}
\end{definition}

The Jordan-Sch\"onflies Theorem (an extension of the Jordan Curve Theorem)
[\ref{stillwell}],
adapted to $S^2$, implies that every Jordan curve $J$ defines a unique
closed disc in $S^2$: the curve may be
oriented in the direction of increasing $\phi$, and
$S^2\backslash J$ is the union of two disjoint open topological
discs of which $J$ is the boundary of both; the one meeting
the oriented curve from its left-hand side is
the interior $D^\circ$ of the disc, and $D=D^\circ \cup J$
is the closed disc.
This gives a way of parametrising closed discs in $S^2$ with differentiable
boundary, by $C^1$ maps.

\numpara
\label{par: C1 metric}
The `sup norm' on parametrisations $f$ of Jordan curves is
\be
\sup_{0 \leq \phi \leq 2\pi} \| f(\phi) \|.
\ee
We use the notation
\be
\|f\|_\infty.
\ee
The $C^1$ norm on $f$ is
\be
\max
\left (
\| f\|_\infty, ~
\left \| \frac{df}{d\phi} \right \|_\infty
\right ).
\ee
\noindent This gives a metric on the space of all
such closed discs in $S^2$.  A {\em compact family of discs}
is a compact set of parametrisations, under this metric.

\subsection{Pre-seams are semi-algebraic}
Suppose that $B_0$ and $B_1$ are disjoint copies of `model' bodies.
The $B_0,B_1$-{\em seam} is the set of points on $\partial B_0$ at which
the tangent plane is also a (supporting) tangent plane to $B_1$.
Since the bodies are rounded, the seam is
homeomorphic to the circle $S^1$ [\ref{hoy}, Lemma 5].

\begin{definition}
\label{def: pre-seam}
Suppose that $B_0 = B^{f_0,a_0} = \{x: ~ f_0(x-a_0) \leq 1\}.$
The $B_0,B_1$ {\em pre-seam}
is the image of the $B_0,B_1$-seam under the outward normal map
$n_{f_0,a_0}: ~ \partial B_0 \to S^2$.
\end{definition}
Recall that the normal map is a homeomorphism
(Proposition \ref{prop: normal homeo}).\

\begin{proposition}
\label{lem: seam is semi-algebraic}
The $B_0,B_1$-seam is semi-algebraic of bounded degree.
{\rm [\ref{hoy}, Lemma 25]}.\qed
\end{proposition}

\begin{corollary}
\label{cor: pre-seam semi-algebraic}
The $B_0,B_1$ pre-seam is semi-algebraic of bounded degree.
\end{corollary}

{\bf Proof.}
Let $B_0 = B^{f,O}$, and recall that there is a finite
covering $S_1, \ldots, S_k$ of $\IR^3$ by semi-algebraic
sets, and for $1 \leq i \leq k$, a polynomial $p_i(x,y,z)$,
such that $f$ agrees with $p_i$ on $S_i$.

Let $S$ be the $B_0, B_1$ seam.  It is semi-algebraic of bounded
degree, and the pre-seam is
\be
n_{f,O} (S)
\ee
Let $P$  be the pre-seam.
Expressed as a union:
\be
P= \bigcup_i n_{f,O} ( S \cap S_i )
\ee
Let us write
\be
P_i = n_{f,O} (S \cap S_i)
\ee
so $P = \bigcup_i P_i$.
It is enough to show that each set $P_i$ is semi-algebraic.

For $\omega$ to be in $P_i$,
\be
\omega^T \omega = 1
\ee
and there exists an $x$, where
\be
x \in S \cap S_i
\ee
and
\begin{gather*}
\omega^T \nabla_{f} (x) > 0 \quad\text{and}\\
(\omega^T \nabla_{f}(x))^2 = \| \nabla_{f} (x) \|^2 .
\end{gather*}
The last equation is derived from the Cauchy-Schwartz
inequality.  But $f$ agrees with the polynomial $p_i(x,y,z)$.
So we can use the following equations to describe a set
of ordered pairs $(\omega, x)$:
\begin{gather*}
\omega^T \omega = 1\quad\text{and}\\
x \in S \cap S_i \quad\text{and}\\
\omega^T \nabla_{p_i} (x) > 0 \quad\text{and}\\
(\omega^T \nabla_{p_i}(x))^2 = \| \nabla_{p_i} (x) \|^2 .
\end{gather*}
This set of pairs is semi-algebraic, and therefore its
projection onto the $\omega$-coordinate is semi-algebraic.
But its projection is $P_i$.\qed

The main fact about pre-seams is that they form compact
families:

\begin{theorem}
\label{thm: seams compact family}
Given a compact family $\cal G$ of convex bodies, the
family $\cal F$ of pre-seams is a compact family of Jordan
curves on $S^2$
(Corollary \ref{cor: pre-seams compact family}).
\end{theorem}

As a consequence we can obtain the stated bounds
on the feature complexity of convex hulls.
This analysis is given as early as possible, and
the lengthy proof of the above theorem
is given last.

\section{Jordan curves in general position, intersection
number, crossing content}
\label{sect: general_position}
We consider a compact family $\cal F$ of $C^1$ Jordan curves on $S^2$.
Our analysis emphasises sets of curves in general position.
Here we show that if curves are not in general position, then
general position can be established by arbitrarily
small perturbations (actually rotations of $S^2$).

\begin{definition}
\label{def: general position}
Two $C^1$ (Jordan) curves are {\em in general position} (relative to each other)
if all intersections are transversal; that is, if $x$ is a point
common to both curves, then the unit tangent vectors (in $\IR^3$) to
those curves are linearly independent.

A list $C_1, \ldots, C_k$ of  (Jordan) curves is in general
position if every two curves from the list are in general
position relative to one another, and no three curves intersect
at the same point.
\end{definition}

\begin{lemma}
\label{lem: rotation closure}
Let $\cal F$ be a compact family of (or rather,
parametrising)  Jordan curves in $S^2$.
$SO(3)$ is the group of all rotations of
$\IR^3$ and of $S^2$. Then the family
\be
\{ f \circ R:~ f \in {\cal F},~ R \in SO(3) \}
\ee
is also compact.
\end{lemma}

{\bf Proof.} Follows directly from the compactness
of ${\cal F} \times SO(3)$, which can be used
to parametrise the extended family.\qed

So we can assume that $\cal F$ is closed under rotations.

\begin{lemma}
\label{lem: perturbation}
Suppose $\cal F$ is closed under rotations.
Then for any $f \in \cal F$ and $\epsilon > 0$,
there exists a copy $g$ of $f$,
such that $\|f-g\|_\infty < \epsilon$ and
all intersections between the  two
curves are transverse.
\end{lemma}

{\bf Proof.}
Let $x,y \in S^2$, $T_x$ and $T_y$ unit vectors tangent
to $S^2$ at $x$ and $y$ respectively.  There exists a
rotation taking $x$ to $y$ and $T_x$ to $T_y$.
For one can easily rotate $x$ to $y$, and follow this by a rotation
around $y$ to align the tangent vectors.

Suppose that $R_1$ and $R_2$ were two such rotations.
Then $R_2^{-1} \circ R_1$ takes $x$ to $x$ and takes
$T_x$ to $T_x$, so it is the indentity map.  In other words,
the rotation is unique.

Suppose $g$ is a copy $f \circ R$ of $f$, where $R$ is some rotation.
If the curves meet non-transversally at any point, then there
exist angles $\phi_1$ and $\phi_2$ such that
\begin{gather*}
f(\phi_1) = g(\phi_2), \quad \text{and either} \\
\frac{df/d\phi_1}{\| df/d\phi_1\|}
=
\frac{dg/d\phi_2}{\| dg/d\phi_2\|},\quad\text{or}\\
\frac{df/d\phi_1}{\| df/d\phi_1\|}
= -
\frac{dg/d\phi_2}{\| dg/d\phi_2\|}.
\end{gather*}

The two choices of sign yield different but almost indistinguishable
cases, so we ignore the second case.  Taking
\begin{gather*}
x = f(\phi_1),\quad
y=g(\phi_2),\\
T_x = \frac{df/d\phi_1}{\| df/d\phi_1\|},\quad\text{and}\quad
T_y = \frac{dg/d\phi_2}{\| dg/d\phi_2\|},
\end{gather*}
we obtain a unique rotation $R_{\phi_1,\phi_2}$
taking $x$ to $y$ and $T_x$ to $T_y$.

Functions in $\cal F$ have domain $[0,2\pi]$; we could
have chosen the domain as $S^1$, but then the notion
of derivative would need elaboration.
But suppose the functions have domain $S^1$; then the
map
\begin{gather*}
S^1 \times S^1 \to SO(3) \\
(\phi_1, \phi_2) \mapsto R_{\phi_1, \phi_2}
\end{gather*}
is a $C^1$ embedding of a $2$-dimensional manifold
into a $3$-dimensional manifold, and hence the
image has measure zero in $SO(3)$ [\ref{gp}, Appendix 1;
\ref{spivak2}, Chapter 2].

Consequently, the set of rotations taking the curve
defined by $f$ to another in general position, is
dense in $SO(3)$.\qed

\begin{lemma}
\label{lem: general position}
Suppose that $\cal F$ is closed under rotations.  Given
Jordan curves $C_1, \ldots, C_k$, there exist arbitrarly
small rotations taking the curves to others in general position.
\end{lemma}

{\bf Sketch proof.}  First deal with non-transverse intersections.
A small rotation of $C_2$ will make all intersections of $C_1$ and
$C_2$ transverse.  Then a small rotation of $C_3$ will extend this
property to $C_1, C_2, C_3$, and so on.  So we arrive, through
arbitrarily small rotations, at a configuration where all intersections
are transverse.  If three curves pass through a point then a small
rotation of one of them will reduce the multiplicity of the
intersection, and so on.\qed

\begin{definition}
\label{def: bounded intersection number}
A family $\cal F$ of $C^1$ Jordan curves parametrising discs
in $S^2$
has {\em bounded intersection number} if for
any pair $D,E$ of discs in general position,
$| \partial D \cap \partial E |$ is bounded.
\end{definition}

Since  pre-seams are semi-algebraic, and any rotated
copies are semi-algebraic,
they have bounded intersection number.
In the introduction, a distinction was made between overlap
and crossway.  It is repeated here.

\begin{definition}
\label{def: overlap and crossway}
Let $D,E$ be two discs in general position (and with
finitely many intersections).  An {\em intersection component}
is a set of the form $\overline{X}$, where $X$ is a connected
component of $D^\circ \cap E^\circ$.  The boundary of any
intersection component can be separated into an even number
$e_1, \ldots, e_k$ of edges, alternately from $D$ and from $E$,
meeting at vertices in $\partial D \cap \partial E$.

An {\em overlap} is an intersection component bounded by two
edges and vertices.  A {\em crossway} is an intersection component bounded
by four or more edges and vertices.
\end{definition}

\begin{definition}
\label{def: positive crossing content}
The natural measure of area on sets in $S^2$ is the
{\em metric measure,} which will be denoted $\mu$.
It has the familiar properties, including
\be
\mu(S^2)  = 4 \pi.
\ee

A family of discs in $S^2$ has {\em positive crossing content}
if there  is a positive lower bound on
\be
\mu (C)
\ee
where $C$ is a crossway between two discs in general position.
\end{definition}

\section{Feature complexity of unions of discs}
\label{sect: feature_c}
This section contains the crucial results about feature
complexity based on bounded intersection number and positive
crossing content.  It is the combinatorial part of the paper;
everything else is geared to proving the necessary
compactness properties.

Based on the following facts
\begin{itemize}
\item
Given a compact family $\cal G$ of convex bodies,
the derived set of pre-seams defines a compact family
of discs in $S^2$ (with bounded intersection number).
\item
If $\cal F$ is a compact family of discs in $S^2$ with
bounded intersection number, then $\cal F$ has positive
crossing content.
\end{itemize}

\noindent and the following

\begin{theorem}
\label{thm: complexity of disc unions}
If $D_1, \ldots, D_n$ is a set of discs in general position
derived from a family $\cal F$ with bounded intersection
number and positive crossing content, then
$\bigcup D_j$ has $O(\lambda_s(dn))$ features, where $s$ and $d$
are constants depending on $\cal F$ and $\lambda_s(m)$ is the
maximum length of an order-$s$ Davenport-Schinzel sequence with
$m$ letters,
\end{theorem}

\numpara
the following theorem is immediate:

{\bf Theorem \ref{thm: main theorem}.}
Supposing that $S = \{B_1, \ldots, B_n\}$ is a disjoint set of
convex bodies derived from $\cal G$, in general position.
As remarked previously,
the feature complexity of $H(S)$ is the sum of the feature
complexities of unions of hidden discs on the bodies, or in
$S^2$; whence $H(S)$ has feature complexity $O(n\lambda_s(dn)$
for constants $s$ and $d$, which is $o(n^2 \log^* n)$.

\noindent (This is Theorem \ref{thm: main theorem}).

This section contains a proof of
Theorem \ref{thm: complexity of disc unions}.
Throughout this section,

\begin{itemize}
\item
$D_1, \ldots, D_n$ is a set of discs in general position derived
from $\cal F$.
\item
If $\bigcup D_n$ has several connected components, then
the total feature complexity is the sum over all components.
Hence we can assume that
\be
\bigcup_{j=1}^n D_j
\ee
is connected.
\item
It follows that every connected component of
\be
S^2 \backslash \bigcup_j D_j
\ee
is {\em simply} connected, since a non-simply-connected
component of the complement would separate different components of
the union.
\item
We fix a constant $\kappa$, a positive lower bound for
the crossing content of $\cal F$.\footnote{We allow for
underestimates in the belief that the exact crossing content
will be almost impossible to calculate.}
\end{itemize}

\begin{definition}
\label{def: hub}
A {\em hub} is either
\begin{itemize}
\item
The closure of a maximal connected union
\be
K_1^\circ \cup \ldots \ldots K_r^\circ
\ee
of interiors of crossways, or
\item
Any disc which contains no crossway.
\end{itemize}
\end{definition}

\begin{lemma}
\label{lem: O(n) hubs}
There are $O(n)$ hubs.
\end{lemma}

{\bf Proof.}
There are at most $n$ hubs which are entire discs.
Given that the crossing
content is $\geq \kappa$,
of the remaining hubs, there are at least as many crossways,
and given that the crossing content is $\geq \kappa$,
and there are at most $4\pi/\kappa$ crossways.\qed

\begin{lemma}
\label{lem: retract overlaps}
For any $\epsilon > 0$, it is possible to
modify the discs $D_j$, without changing their
external boundary (i.e., without changing the features
of $\bigcup D_j$),
so that for every disc $D_i$, all overlaps between $D_i$
and other discs are contained in $N_\epsilon ( \partial D_i )$.
(Proof omitted.)\qed
\end{lemma}

\begin{figure}
\centerline{\includegraphics[height=1in]{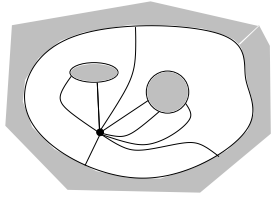}}
\caption{connecting the `centre' to points on $\partial C$.}
\label{fig: linkup}
\end{figure}

\begin{lemma}
\label{lem: radiating}
Let $C$ be a closed subset of $S^2$ whose boundary is
a finite union of closed (piecewise $C^1$) loops. Choose
a point $c$ in $C$, called  a `centre.'  Then, given
a finite set $x_1, \ldots, x_t$ of point in $\partial C$,
one can connect $c$ to all points $x_j$ by simple continuous paths
which are pairwise disjoint except where they meet at $c$.
(Proof omitted; see Figure \ref{fig: linkup}).\qed
\end{lemma}

\begin{lemma}
\label{lem: O(n) overlaps}
Overlaps contribute $O(n)$ (external) vertices to
$\bigcup D_j$.
\end{lemma}

{\bf Proof.} 
Apply Lemma \ref{lem: retract overlaps} with an $\epsilon$
sufficiently small so that for every hub $K$,
$K \backslash  N_\epsilon ( \partial K )$ is connected,
and then choose a centre in each hub.

For each pair $D_i, D_j$ which share a boundary vertex
on an overlap $V$, choose
one, $V$, of these overlaps, and an external vertex $v$ on $V$.
The point $v$ is in $D_i \cap D_j$ and disjoint from
all other discs.

Choose points $x\in D_i$ and $y \in D_j$
on the interior of their bounding  edges and sufficiently
close to $v$ so that they are contained in no discs
except $D_i$ and $D_j$.

If the disc $D_i$ intersects a crossway, then one can form a path
which joins $x$ to a point $x'$ in a hub boundary in $D_i$.
Applying Lemma \ref{lem: radiating}, $x$ can be joined
to the centre of the hub.  Otherwise $D_i$ is itself a
hub and one can join $x$ to its centre.  Similarly for $y$.

This defines a planar graph whose edges join the centres of hubs
and which therefore has $O(n)$ edges.  Therefore there are
$O(n)$ pairs $D_i, D_j$ which can intersect in an overlap (which
meets $\partial \bigcup D_\ell$).
Each such pair possesses $O(1)$ overlaps by bounded intersection
number. Hence there are $O(n)$ such overlaps.\qed

\begin{definition}
\label{def: link}
Suppose that $D_i$ is a disc and $U$ is the union of crossways:
by the arguments in Lemma \ref{lem: O(n) hubs},
$D_i \cap U$ has $O(1)$ components.
Different components may be part of the same hub.
Let $K_1, \ldots, K_k$ be these components. In Figure
\ref{fig: link} they are shaded.  (Only crossways are
considered here; possible overlaps are omitted from the figure.)

\begin{figure}
\centerline{\includegraphics[width=1.5in]{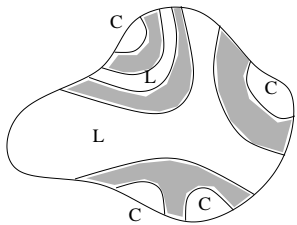}}
\caption{\em connected unions of crossways in $D_i$, links,
and coves. Overlaps are ignored.}
\label{fig: link}
\end{figure}
\be
X = D_i \backslash K_1 \ldots \backslash K_k
\ee
has a potentially unbounded number of components,
but if we distinguish {\em links} from {\em coves}
there is a bounded number of links.

A {\em link} in $D_i$ is either $D_i$ itself, if $k=0$
($D_i$ has no crossways), or it
is (the closure of) a component of the
above subset $X$ of $D_i$ whose intersection with
$\partial D_i$ is nonempty and disconnected.

A {\em cove} is
a component whose intersection with $\partial D_i$
is nonempty and connected.

An {\em external link segment} is a connected component of
$L \cap \partial D_i$, where  $L$ is a link in $D_i$.
\end{definition}

\begin{lemma}
\label{lem: O(1) links}
In each disc $D_i$ there are $O(1)$ links and external link
segments.\footnote{There can be arbitrarily many coves.}
\end{lemma}
\noindent {\bf Proof.}
Choose any link $L$. $D_i \backslash L$ is disconnected,
Let $R_1,\ldots, R_\ell$ be the closures of the
components of $D_i\backslash L$.

Every component $K_j$ of $D_i  \cap U$ is contained
in one of the $R_j$, so this partitions $K_1,\ldots, K_k$
into $\ell \geq 2$ groups.  Continuing in the same way
with the sets $R_1,\ldots, R_\ell$, recursively, we obtain
a recursive partition of $K_1,\ldots, K_k$, a tree
structure in which every internal node has degree $\geq 2$,
and in which every leaf carries one of the components $K_j$.

The tree has fewer than $k$ internal nodes.  Each internal
node corresponds to a link, and the number of children it has
matches the number of external link segments. Thus there are
$O(k)$ links and $O(k)$ external link segments, and $k$
is bounded.\qed

We have established that there are $O(n)$ overlaps, or at least
that $O(n)$ overlaps can contribute features to $\bigcup D_j$.
In the proof, we imagined shrinking the overlaps --- retracting
them --- and using planarity arguments.  We can retract the overlaps
still further, and obtain the following result:

\begin{lemma}
\label{lem: further retraction}
The overlaps can be retracted further
so they disappear, removing $O(n)$ features from $\bigcup D_j$, without
adding or removing any other vertices.
This leaves a union of discs which intersect only at crossways, with
no overlaps.\qed
\end{lemma}

\begin{definition}
\label{def: hole}
A {\em hole} is the closure of a connected component
of $S^2 \backslash \bigcup D_i$.

Since the union is assumed connected, every hole is simply connected.
\end{definition}

\begin{lemma}
\label{lem: partial unions connected}
Combinatorial lemma: if  $\bigcup_1^n D_i$ is connected, then by re-ordering the list
$D_1,\ldots, D_n$ if necessary, it can be arranged that
every partial union
$\bigcup_1^k D_i$, $1 \leq k \leq n$, is connected.
\end{lemma}

{\bf Proof.} Form the intersection graph $G$
whose vertices are $\{1,\ldots, n\}$
and edges are $\{\{ i,j \}: D_i \cap D_j \not= \emptyset\}$.
For any set $S$ of vertices, the corresponding union of discs is connected
if and only if the subgraph spanned by $S$ is connected.

Given that $G$ is nonempty, it contains a vertex $v$ which is
{\em not} an articulation point, as follows.  Let $T$ be a spanning
tree for $G$.  Let $v$ be a leaf of $T$.  Then $T \backslash \{ v\}$ is
also a tree, and therefore $G \backslash \{v\}$ is connected, as claimed.

Let $G_n = G$.  Choose a vertex $v_n$ which is not an articulation point.
Let $G_{n-1} = G \backslash \{v_n\}$.  By induction on $n$ we can assume
that $G_{n-1}$ has the stated property, and therefore so has $G$.\qed

\begin{lemma}
\label{lem: disc-hole pairs}
There are $O(n)$ pairs $D_i,H_j$ where $H_j$ is a hole
incident to $D_i$.
\end{lemma}

\noindent {\bf Proof.}
We can assume that $\bigcup_1^k D_i$ is connected
for $1\leq k \leq n$.  We apply induction on $k$.
Suppose the disc $D_k$
is added to an existing union $\bigcup_1^{k-1} D_i$ ($k \geq 2$).
It is enough to show that $O(1)$ new holes are created.

The number of holes is increased by virtue of an existing
hole, or holes, $H$, being split into several, $H_1,\ldots, H_\ell$, by
$D_k$.  The holes are always simply connected.

Let $H_r$ and $H_s$ be holes, part of the same hole $H$
split by $D_k$.  $H$ is (simply) connected.  Consider any
path in $H$ joining points $y_r$ and $y_s$ interior to $H_r$ and $H_s$.
The path crosses $\partial D_k$ at least twice.  If the path
wanders into a cove from $H_r$, it must wander out again without
leaving $H_r$.  So the path must cross some external link segment
incident to $H_r$.  Thus all the holes $H_r$ are incident to
external link segments in $D_k$: there are $O(1)$ external link segments,
so adding the disc $D_k$ creates $O(1)$ new holes.\qed

\begin{corollary}
\label{cor: bigcup Di has feature complexity}
There exist constants $s$ and $d$ such that
$\bigcup D_i$ has feature complexity
$O( \lambda_s ( dn ))$.
\end{corollary}

\noindent {\bf Proof.}
For any $H_i$, suppose there are $d_i$ discs $D_j$ sharing an
edge with $H_i$; $\sum_i d_i \leq dn$ for some constant $d$.

Let $e_1,\ldots, e_k$ be the edges incident
to $H_i$, in anticlockwise order; each edge is on one of the
discs $D_j$, and may be labelled with the index $j$. We
get a list $j_1, \ldots, j_k$ of indexes.  Of course no
index is repeated twice in succession, and since the discs
have bounded intersection number, there is an upper bound
$s$ on the length of alternating subsequences.

We have a Davenport-Schinzel sequence.
Therefore for some constant $s$, $H_i$ has
$\leq \lambda_s (d_i)$ edges. Adding,
$\bigcup D_j$ has $O(\lambda_s ( dn ))$ features.\qed

\section{Compact families of discs have positive crossing content}
\label{sect: positive crossing content}
Recall the definition of positive crossing content
(\ref{def: positive crossing content}).
In this section we prove:

\begin{theorem}
\label{thm: positive crossing content}
If $\cal F$ is a compact family of Jordan curves
in $S^2$ (with bounded intersection number), then 
$\cal F$ has positive crossing content, i.e.,
\be
\takeanumber
\label{eq: crossing content}
\tag{\thetheorem}
\inf_K \mu(K) > 0,
\ee
where $K$ ranges over all crossways from pairs
of discs in general position.
\end{theorem}

Actually, our proof makes little reference to measure theory; it just uses
the following elementary fact.

\begin{proposition}
\label{prop: measure nonempty interior}
If $K \subseteq S^2$ is closed (therefore
measurable) and $K^\circ \not= \emptyset$,
then $\mu(K) > 0$.\qed
\end{proposition}

{\bf Proof strategy.}
The general idea is that if $K_n$ is a sequence of
crossways, convergent, in an informal sense\footnote{Our
proof does not mention Hausdorff distance.} to a set
$K$, then $K$ is bounded by a well-defined sequence
of edges, and if they do not enclose any open set then
the angle between successive edges is $360^\circ$, which
is only possible if there are two edges and the sets $K_n$ are
overlaps: a contradiction.

\begin{definition}
\label{def: widening}
\be
\overline{W_\epsilon} = S^2 \cap \overline{N_\epsilon}(S)
=
\{ x \in S^2:~ (\exists y \in S)~ \| x-y \| \leq \epsilon \}.
\ee
\end{definition}
We call $\overline{W_\epsilon}$ the
{\em (closed) $\epsilon$-widening} of $S$.
A related idea of `thickening' in three dimensions
will be introduced in Definition \ref{def: thickening}.

\begin{lemma}
\label{lem: widening contains curve}
If $D_1$ and $D_2$ are discs bounded by Jordan curves parametrised by
functions $f_1$ and $f_2$, and $\| f_1 - f_2 \|_\infty \leq \epsilon$,
then each boundary is within the $\epsilon$-widening
of the other. (Trivial.)\qed
\end{lemma}

\begin{lemma}
\label{lem: nbd and boundary of J curve semi-algebraic}
Suppose a Jordan curve in $S^2$ is
a semi-algebraic subset $S$ of $S^2$. Given $\epsilon > 0$,
the widening $\overline{W_\epsilon}$
is semi-algebraic.
Also, its  interior (replacing `$\leq$' by `$<$' in the definition) and its
boundary are semi-algebraic.

Also, its inner boundary: let $D$ be the closed disc to
the left of $S$, when the orientation of $S$ is taken into
account.  The inner boundary is
\be
D \cap \partial  \overline{W_\epsilon}.
\ee
Likewise, the outer boundary $\partial \overline{W_\epsilon} \backslash D$.
(See {\rm [\ref{hoy}, Theorem 3]},
or {\rm [\ref{bpr}]}).\qed
\end{lemma}

These boundaries may be self-intersecting in the sense
that as algebraic curves they have double points.  We need
to bound the number of double points.

\begin{lemma}
\label{lem: double points}
{\rm (i)} Let $Y = \{(x,y,z)\in S^2: ~ p(x,y,z) = 0 \}$
be an algebraic curve where $p$ is a polynomial of degree $k$.
Then $Y$ contains at most $k^2$ double points.

More generally, {\rm (ii)} if $Y$ is semi-algebraic of bounded degree,
then $Y$ has a bounded number of double points.
\end{lemma}

{\bf Sketch proof.} (i) Let $p\circ R$ be a copy of $p$ obtained
by a small rotation, defining
a curve $Z$, so the two curves are in general position.
If $R$ is sufficiently small, then every double point of $Y$
is close to two intersection points of $Y \cap Z$, and
there are at most $k^2$ such intersections by B\'{e}zout's Theorem.
This is easily generalised to obtain (ii).\qed

\begin{figure}
\centerline{\includegraphics[height=1in]{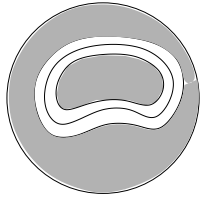}}
\caption{$S^2 \backslash \overline{W_\epsilon}$ has two (simply)
connected components.}
\label{fig: strip}
\end{figure}

\begin{corollary}
\label{cor: connected inside widening}
Given a closed disc $D$ in $S^2$
whose boundary is an oriented $C^1$ semi-algebraic Jordan curve,
then for sufficiently small $\epsilon$,
$S^2\backslash \overline{W_\epsilon}$ is the union
of two open simply-connected regions, one inside
$D$ and one outside.
\end{corollary}

\noindent{\bf Proof.}
See Figure \ref{fig: strip}.
First, if $\epsilon$ is small enough, then
$D \backslash \overline{W_\epsilon} \not= \emptyset$:
choose any point $x \in D^\circ$, let $\delta = d(x, \partial D)$,
let $\epsilon = \delta/3$, and let $C = S^2 \cap \overline{N_\epsilon(x)}$.
Then $d(C, \overline{W_\epsilon}) = \delta/3$, and
$C \subseteq D\backslash \overline{W_\epsilon}$.

For all sufficiently small $\epsilon$,
$D\backslash \overline{W}$ is nonempty and contains
a finite number of connected components, by
Lemma \ref{lem: double points}.
As $\epsilon$ decreases, the components grow
and coalesce. Since there are finitely many components,
the number of components must reach a minimum $m$.

If $m>1$, there would exist
two points $x$ and $y$ in $D^\circ$ which belong to different components
of $D\backslash W_\epsilon$ for all sufficiently small
$\epsilon$.  Let $P$ be any path from $x$ to $y$ in $D^\circ$,
and let $0 < \delta <d(P, \partial D)$.  Then $P$
is a path from $x$ to $y$ in $D\backslash \overline{W_{\delta}}$,
a contradiction.\qed

This says that Jordan curves are not `pinched.'  That can
also be viewed as relating Euclidean distance to distance
along the curve.

\begin{definition}
Let $[a,b]$ be a closed interval and $c:\, [a,b]\to S^2$;
$t \mapsto c(t)$ be a $C^1$ parametrisation of a (not necessarily simple)
curve $C$.
The {\em length} $\lambda(C)$ of $C$ is the limit
as $n\to\infty$ of
\be
\sum_{i=0}^k \| c(t_{i+1}) - c(t_i) \|
\ee
where $t_0, t_1, \ldots, t_k$ is a division of
the interval $[a,b]$ into even-width intervals.
Put differently:
\be
\takeanumber
\tag{\thetheorem}
\label{eq: length of curve}
\lambda (C) = \int_a^b \left \| \frac{dc}{dt} \right \| dt.
\ee
\end{definition}

\begin{lemma}
\label{lem: length continuous}
If $C$ and $\hat{C}$ are $C^1$ curves with nearby
parametrisations under the
$C^1$ metric, then $|\lambda(C)-\lambda(\hat{C})|$ is small.
\end{lemma}

\noindent{\bf Sketch proof.} We are comparing something
like (\ref{eq: length of curve}) with something like
\be
\lambda (\hat{C}) = \int_{\hat{a}}^{\hat{b}}
\left \| \frac{d\hat{c}}{dt} \right \| dt.
\ee
where $a$ is close to $\hat{a}$, $b$ to $\hat{b}$, and
$c$ is close to $\hat{c}$ on their common domain of definition,
and the derivatives are also close.

In that case, $\| dc/dt\|$ and $\| d\hat{c}/dt\|$ are close
and the lengths are close.\qed

\begin{lemma}
\label{lem: no pinch}
Let $D$ be a disc (in $S^2$ with oriented $C^1$ boundary).
$\partial D$ is rectifiable.
Given
points $x,y \in \partial D$, let $\lambda(x,y)$ be the distance
along $\partial D$ (anticlockwise) from $x$ to $y$  and let
\be
\rho(x,y) = \min(\lambda(x,y), \lambda(y,x)).
\ee
Then for all sufficiently small $\epsilon>0$, there exists a
$\delta>0$ such that for all $x,y \in \partial D$,
\be
\| x-y \| < \epsilon \implies \rho(x,y) < \delta.
\ee
\end{lemma}

{\bf Proof.}
Otherwise there exists a $\delta > 0$ and a sequence
of pairs $x_n, y_n$ in $\partial D$ such that $\| x_n - y_n\|\to 0$
and $\rho(x_n,y_n) \geq \delta$.  Since $\partial D$ is compact
we may choose a subsequence if necessary so $x_n$ and $y_n$ both
converge to points $x$ and $y$ respectively in $\partial D$.

\begin{figure}
\centerline{\includegraphics[height=1in]{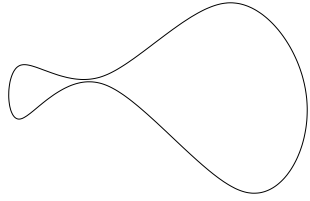}}
\caption{distance along $\partial D$ doesn't outstrip Euclidean
distance on $S^2$.}
\label{fig: pinch}
\end{figure}

But then $x=y$ and there are two closed subpaths of $\partial D$
meeting at $x$, so $\partial D$ would not be a Jordan curve.
See Figure \ref{fig: pinch}.\qed

\begin{figure}
\centerline{\includegraphics[height=1in]{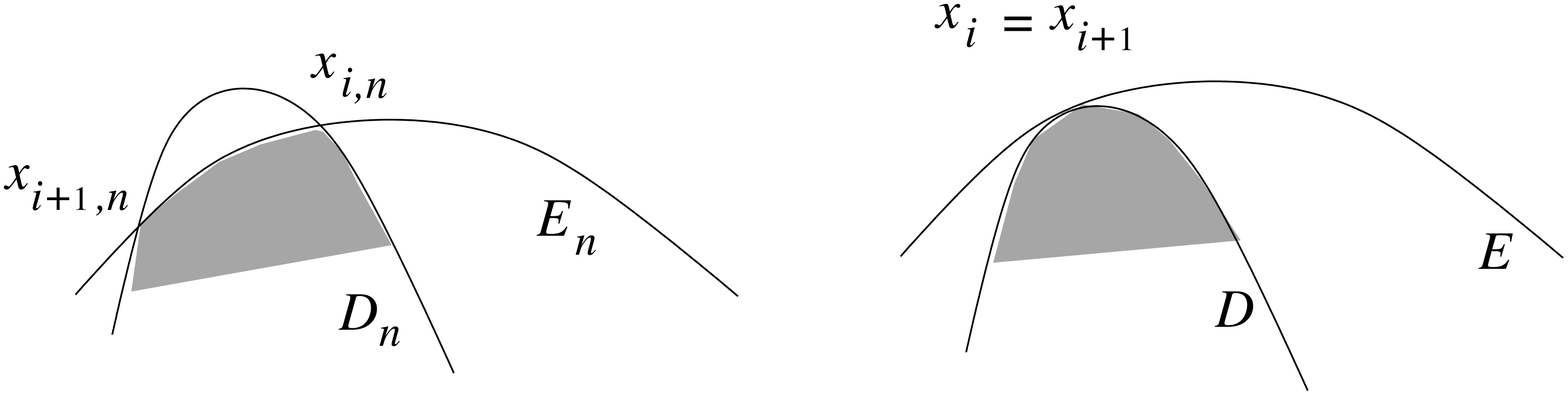}}
\caption{$x_i \not= x_{i+1}$.}
\label{fig: poscc_nondegen}
\end{figure}

\begin{figure}
\centerline{\includegraphics[height=1in]{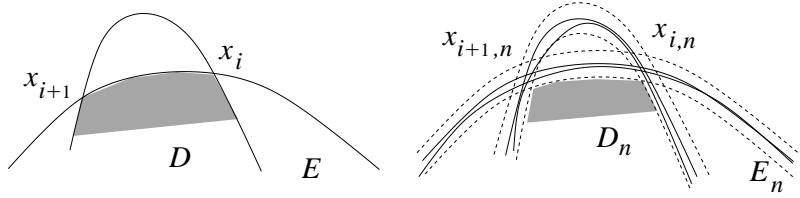}}
\caption{All limiting edges are in $\partial D \cap \partial E$.}
\label{fig: poscc_side}
\end{figure}

\hb

We begin the proof of positive crossing content as follows.  Suppose that
(\ref{eq: crossing content}) is false.  Then for every
$\epsilon > 0$ there exists a crossway $K$ whose
measure is $\leq \epsilon$. Therefore there exists
a sequence
\be
K_n
\ee
of crossways such that $\mu (K_n) \to 0$.
Since $\cal F$ has bounded intersection number,
there exists an even integer $k \geq 4$ such that
infinitely many of these crossways $K_n$ have
$k$ edges.

Fix such a $k$ and discard the  other terms in the sequence.
Now each of these crossways can be defined by a tuple
\be
D_n , E_n , x_{1n}, \ldots , x_{kn}
\ee
where $x_{jn}$ are the vertices of $K_n$ in anticlockwise
order, and (for definiteness) the edge joining $x_{1n}$ to
$x_{2n}$ around $K_n$ is part of $\partial D_n$.

Since $\cal F$ is compact and $S^2$ is compact, we may assume
that these tuples converge to a limit
\be
D, E, x_1, \ldots, x_k .
\ee

The discs $D$ and $E$ are probably not in general position,
but there is a well-defined sequence of edges
$e_1, e_2, \ldots, e_k$
joining $x_1 $ to $x_2$ in $\partial D$,
$x_2 $ to $x_3$ in $\partial E$, and so on.

Clearly the union of these edges is connected.

\begin{lemma}
For $1\leq j < k$ the vertices $x_j$ and
$x_{j+1}$ ($x_{k+1} = x_1$) are distinct. For otherwise
there would be an open region in $D\cap E$ to their left,
containing an open regions $R$ such
that for all sufficiently large $n$,
$R \subseteq K_n$, so $\mu(K_n)\geq \mu(R) > 0$.
See Figure \ref{fig: poscc_nondegen}.\qed
\end{lemma}

\begin{lemma}
These edges are all in $\partial D \cap \partial E$.
Otherwise let $e$ be an edge from $\partial E$, say, which
intersects $D^\circ$.  Then to the left of the edge there
is an open nonempty subset $X$ of $D \cap E$.
There would be an $\epsilon > 0$ with the property
that $X \backslash N_\epsilon( \partial D) \backslash
N_\epsilon(\partial E)$ would have positive measure,
and it would follow that $\mu (K_n)$ is bounded below.
See Figure \ref{fig: poscc_side}.\qed
\end{lemma}

\begin{figure}
\centerline{\includegraphics[height=1.5in]{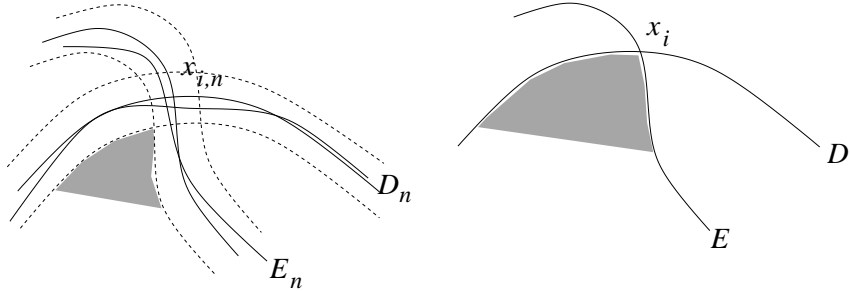}}
\caption{Angle at $x_i$ must be $360^\circ$.}
\label{fig: poscc_corner}
\end{figure}

\begin{lemma}
For all sufficiently large $n$, at all corners of $K_n$ the
tangents meet at reflex angles. See Figure
\ref{fig: poscc_corner}.\qed
\end{lemma}

\noindent
{\bf Sketch proof} of Theorem
\ref{thm: positive crossing content}.

All limiting edges are nondegenerate and contained in
$\partial D \cap \partial E$. Consider the edge
$e_1$ joining $x_1$ to $x_2$, which is (without loss
of generality, or by assumption) a limit of edges
$e_{1n}$ belonging to $\partial D_n$.
Let $e'$ be the other edge incident to $x_2$.
Then $e'$ is the limit of edges in $E_n$.

Claim that $e'$ joins $x_2$ to $x_1$.

Suppose otherwise: $e'$ joins $x_2$ to another vertex
$y$. The angle at $x_2$ is reflex ($360^\circ$)
so $x_1$ and $y$ are on the same side of $x_2$ in
$\partial D \cap \partial E$.  Either $y$ is between
$x_1$ and $x_2$ or $x_1$ is between $y$ and $x_2$.
These cases are much the same: assume the first.

\begin{figure}
\centerline{\includegraphics[height=2in]{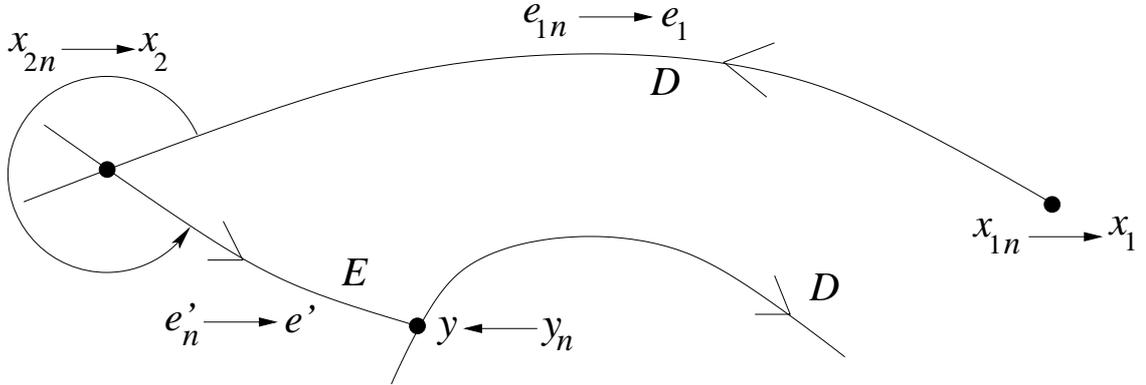}}
\caption{two edges join $x_1, x_2$.}
\label{fig: doubleback}
\end{figure}

The edge $e'$ is a limit of edges $e'_{1n} \subseteq E_n$,
and $e_1$ is a limit of edges $e_{1n} \subseteq D_n$.  See
Figure \ref{fig: doubleback}.

Now the other endpoint $y$ of $e'$ is a limit of endpoints
$y_n$. For each $n$, let $z_n$ be a point in $e_{1n}$ closest
to $y_n$.  Then $\| y_n - z_n \| \to 0$.  But their separation
along $\partial D_n$ is bounded below by
\be
\min ( \|y_n - x_{1n}\|, \| y_n - x_{2n} \| )
\ee
and this distance is bounded below, in the limit, by
\be
\min ( \| y-x_1\|, \| y-x_2\| )
\ee
which is positive, contradicting Lemma \ref{lem: no pinch}.

Therefore $e'$ joins $x_2$ to $x_1$, $e' = e_2$, and for
large $n$ the corresponding edges $e_{1n}, e_{2n}$
surround a connected
component of $D_n \cap E_n$, i.e., $K_n$: so 
$K_n$ is an overlap, not a crossway, a contradiction.\qed

\section{Pre-seams are differentiable Jordan curves with
bounded intersection number}
\label{sect: pre-seams are differentiable}
{\bf Preview.}
It is asssumed in this section that all bodies mentioned
are derived from a compact family $\cal G$.

This section includes some important material about the
continuity of the map $f,a \mapsto n_{f,a}$, the latter
being the outward normal map. It introduces the important
notion of a {\em thickening} (of $\partial B$),
and it introduces the important
notion of a {\em pair descriptor}, which is necessary to define
the map from pairs $B_0, B_1$ to the pre-seams.
Then it is proved that the pre-seam is a $C^1$ manifold,
by a routine application of the Implicit Function Theorem.
Next it is shown that the pre-seam, under its standard parametrisation,
is a $C^1$ map.  The section concludes with a brief proof that
the pre-seams are semi-algebraic of bounded degree.

\begin{definition}
\label{def: f, a, normal, p}
Given
\be
B= \{ x: ~ f(x-a) \leq 1\}
\ee
and $n$ is its normal map (at or near the boundary $\{x:~f(x-a)=1\}$),
\be
n(x) = n_{f,a}(x) =  \frac{\nabla_f(x-a)}{\|\nabla_f(x-a)\|},
\ee
(the subscripts $f,a$ may be omitted if they are clear from the context).
We define a right inverse to
$n$, $p_{f,a}: \IR^3 \backslash\{O\} \to \partial B$:
\be
p_{f,a} ( y ) = n^{-1}
\left ( \frac{y}{\| y \|}\right ).
\ee
\end{definition}
\noindent
The map $p_{f,a}$
is well-defined and continuous
because $n$ is a homeomorphism from $\partial B$
onto $S^2$ [\ref{hoy}, Lemma 1].
Recall (Section \ref{subsec: neighbourhoods}) that 
\be
N_\epsilon(\ldots)
\ee
denotes
an open $\epsilon$-neighbourhood as understood for objects
of various kinds under various metrics.

\begin{definition}
\label{def: thickening}

Given $B = B^{f,a}$, we write
\be
\overline{\Theta_\epsilon}
\ee
for
\begin{gather*}
\overline{N_\epsilon(\partial B)} =
\{ x \in \IR^3:~ d(x, \partial B) \leq \epsilon\} =
\{ x \in \IR^3:~ \inf_{y\in \partial B} \| x-y \| \leq \epsilon\} .
\end{gather*}
We call $\overline{\Theta_\epsilon}$
the {\em closed $\epsilon$-thickening} of $\partial B$.
Its dependence on $f$ and $a$, and thus $B$, is left implicit.
\end{definition}

\begin{lemma}
\label{lem: shell thickening}
Given $\hat{B} = B^{\hat{f},\hat{a}}$, and $\epsilon > 0$,
let $\overline{\Theta_\epsilon}$ be the $\epsilon$-thickening
of $\partial \hat{B}$. Then there exists a neighbourhood
$U$ of $\hat{f},\hat{a}$, under the product metric
on ${\cal G}\times \IR^3$, such that
\be
\forall f,a \in U ~~~
\partial B^{f,a} \subseteq \overline{{\Theta}_\epsilon}.
\ee
\end{lemma}

\noindent{\bf Proof.} Without loss of generality, $\hat{a} = O$.
First we consider the simpler case where $a = O$.

\noindent
Since $O \in \hat{B}^\circ$, we can assume without loss
of generality that $d(O,\partial \hat{B}) \geq \epsilon$.
Let
\be
I = \{ x \in \hat{B}: d (x, \partial \hat{B} ) \geq \epsilon\} .
\ee
Note $I \not=\emptyset$.
For all points $x\in I$, $\hat{f}(x) < 1$; also,
$I$ is compact, so there exists $\delta > 0$ such that
\be
\takeanumber
\label{eq: 1 minus delta}
\tag{\thetheorem}
(\forall x \in I ) \quad  \hat{f}(x) < 1 - \delta.
\ee

\noindent
Recall $B^{f,O} \subseteq \overline{N_{3/2}(O)}$ for all $f \in \cal G$,
because $f(x) \equiv 2$ outside $N_{3/2}(O)$.

\noindent
Let
\be
J = \{ x \in \overline{N_2(O)}:~ d(x,\hat{B}) \geq \epsilon\}.
\ee
Without loss of generality, $\epsilon \leq 1/2$: $\partial N_2(O)$
is the sphere of radius $2$, and, since $\hat{B} \subseteq N_{3/2}(O)$,
$\partial N_2(O) \subseteq J$.
Note
\be
\overline{N_2(O)} \backslash ( I \cup J ) \subseteq
\overline{{\Theta}_{\epsilon}}.
\ee

\noindent
Revise the above choice of $\delta$ (Equation \ref{eq: 1 minus delta}) so that
\be
(\forall x \in J) \quad f(x) > 1 + \delta.
\ee

\noindent
For any $f \in N_\delta (\hat{f})$, (the $\delta$-ball in the
$C^2$ metric),
if $x \in I$, then
\be
f(x) = f(x) - \hat{f}(x) + \hat{f}(x) < \delta + 1 - \delta = 1,
\ee
so $x \in (B^{f,O})^\circ$.
If $x \in J$, then
\be
f(x) = f(x) - \hat{f}(x) + \hat{f}(x) > - \delta + 1 + \delta = 1,
\ee
so $x \in \IR^3 \backslash \overline{B^{f,O}}$. Therefore, if
$x \in \partial B^{f,O)}$ then $x \notin I$ and $x \notin J$,
so $x \in \overline{{\Theta}_\epsilon}$.

To finish the result, choose $\delta > 0$ so that
for all $f$ in $N_\delta \hat f$,
$\partial B^{f,O} \subseteq \overline{{\Theta}_{\epsilon/2}}$.

\noindent
Now to define the neighbourhood $U$ of $\hat{f},\hat{a}$ ($\hat{a} = O$):
\be
U = N_\delta(\hat {f}) \times N_{\epsilon/2}(O) .
\ee
Then for all $f,a \in U$
\be
\partial B^{f,a} = 
a + \partial B^{f,O}
\subseteq
a + \overline{{\Theta}_{\epsilon/2}}
\subseteq
\overline{{\Theta}_{\epsilon}} . \qed
\ee

\begin{lemma}
\label{lem: thickening n f a - hatted < epsilon}
Given $\hat{B} = B^{\hat{f},\hat{a}}$, and $\epsilon>0$, there exists
a $\delta$-thickening $\overline{{\Theta}_\delta}$ of
$\partial \hat{B}$, and a neighbourhood $U$ of $\hat{f},\hat{a}$
such that for all $f,a \in U$, $\partial B^{f,a} \subseteq 
\overline{{\Theta}_\delta}$ and the outer normal $n_{f,a}$
is defined throughout $\overline{{\Theta}_\delta}$, and
$\|n_{f,a} - n_{\hat{f},\hat{a}} \| < \epsilon$ (or,
equivalently, $\leq \epsilon$)  uniformly
throughout $\overline{{\Theta}_{\delta}}$.
\end{lemma}

\noindent {\bf Proof.}
Without loss of generality, $\hat{a} = O$.

Since
$\nabla_{\hat{f}}(x)$ is nonzero on
$\partial \hat{B}$, we can
choose $\delta>0$ so that
\begin{gather*}
\nabla_{\hat{f}}(x)  \not= O \quad\text{on}\quad
\overline{{\Theta}_{2\delta}},
\end{gather*}
and $n_{\hat{f},O}$ varies by $\leq \epsilon/2$ on
$\overline{{\Theta}_{2\delta}}$: i.e.,

\centerline{
for
all $x,y \in 
\overline{{\Theta}_{2\delta}}$,
$\|n_{\hat{f},O}(x) - n_{\hat{f},O}(y) \| \leq \epsilon/2.$
}
\noindent Let
\begin{gather*}
m = \inf \{\|\nabla{\hat{f}}(x)\|: ~
x \in  \overline{{\Theta}_{2\delta}}\} .
\end{gather*}
Choose  $\eta > 0$ so that, firstly, for all $f \in \cal G$,
if $\|f-\hat{f}\|_{C^2} < \eta$, then
\begin{gather*}
\| \nabla_f(x) - \nabla_{\hat{f}}(x) \|_\infty
\leq \frac{m\epsilon}{8}.\\
\end{gather*}
Note that for all such $f$ and $x$, where $x \in \overline{\Theta_{2\delta}}$,
using the triangle inequality,
\be
\| \nabla_f(x) \| \geq m - \frac{m\epsilon}{8},
\ee
and, assuming $\epsilon \leq 4$,
\be
\| \nabla_f(x) \| \geq \frac{m}{2}.
\ee
The second requirement for $\eta$ is that for all
$f \in N_\eta(\hat{f})$,
\be
\partial B^{f,O} \subseteq \overline{{\Theta}_\delta}
\ee
(note: $\delta$, not $2\delta$).

The set $U$ will be $N_\eta(\hat{f}) \times N_\delta(O)$.
We shall show presently that for every $f \in N_\eta(\hat{f})$,
\be
\takeanumber
\label{eq: n f,O - n hat f,O < epsilon/2}
\tag{\thetheorem}
\sup\{\| n_{f,O}(x) - n_{\hat{f},O}(x) \|: ~
x \in \overline{{\Theta}_{2\delta}} \}\quad\leq \quad
\frac{\epsilon}{2}.
\ee
Then for every $f \in N_\eta(\hat{f})$, $ a  \in N_{\delta}(O)$,
and $x\in \overline{{\Theta}_\delta}$, noting
that $x-a \in \overline{{\Theta}_{2\delta}}$,
\begin{gather*}
\| n_{f,a}(x) - n_{\hat{f},O}(x) \| \leq \\
\| n_{f,a}(x) - n_{\hat{f},a} (x) \| +
\| n_{\hat{f},a}(x) - n_{\hat{f},O}(x) \| = \\
\| n_{f,O}(x-a) - n_{\hat{f},O}(x-a) \| +
\| n_{\hat{f},O}(x-a) - n_{\hat{f},O}(x) \| \leq \\
\frac{\epsilon}{2} + \frac{\epsilon}{2} = \epsilon.
\end{gather*}

\noindent
It remains to prove the inequality
(\ref{eq: n f,O - n hat f,O < epsilon/2}).

Given $f$, write $g(x)$ for $\nabla_{f}(x)$,
$n(x)$ for $g(x)/\|g(x)\|$,
$\hat{n}$ and $\hat{g}$ similarly ($\hat{f}$ in place of $f$).
Given $ x\in \overline{{\Theta}_{2\delta}}$,
\begin{gather*}
n(x) - \hat{n}(x) =
\frac{g(x)}{\|g(x)\|}
-
\frac{\hat{g}(x)}{\|\hat{g}(x)\|} =
\frac{g(x)-\hat{g}(x)}{\|g(x)\|}
+
\hat{g}(x)
\left (
\frac{1}{\|g(x)\|}
-\frac{1}{\|\hat{g}(x)\|}
\right ).
\end{gather*}
Since
\be
\frac{\|g(x) - \hat{g}(x)\|}{\|g(x)\|}
\leq 
\frac{ m\epsilon /8}{m/2} \leq \epsilon/4
\ee
and
\be
\| \hat{g}(x) \|
\left |
\frac{1}{\|g(x)\|}
-\frac{1}{\|\hat{g}(x)\|}
\right |
=
\left | \frac{\|\hat{g}(x) \|}{\|g(x)\|} - 1 \right | \leq\\
\left |
\frac{\|\hat{g}(x)-g(x)\|}{\|g(x)\|} + \frac{\|g(x)\|}{\|g(x)\|}-1
\right |
\leq \frac{m \epsilon/8}{m/2} = \epsilon/4,
\ee
$\| n(x)-\hat{n}(x) \| \leq \epsilon/2$, as required.\qed

\begin{corollary}
\label{cor: p sub f,a continuous}
Let $C(S^2)$ be the space of continuous functions
from $S^2$ to $\IR^3$. There is a natural metric on $C(S^2)$,
\be
\| g_1 - g_2 \|_{S^2} = \sup_{\omega\in S^2} \|g_1(\omega)-g_2(\omega)\|.
\ee

The map $f,a \mapsto p_{f,a}$ (Definition \ref{def: f, a, normal, p})
is continuous from 
the $C^2 \times \|\ldots\|$ metric to $C(S^2)$ under
this metric (on the restriction of $p_{f,a}$ to $S^2$).
\end{corollary}

\noindent{\bf Proof.}
Given $\hat{f},\hat{a}$, and  $\epsilon > 0$, we want a neighbourhood $U$ of
this pair so that for every $f,a \in U$ and $\omega \in S^2$,
\be
\| p_{f,a}(\omega) - p_{\hat{f},\hat{a}}(\omega) \| < \epsilon.
\ee

\noindent
First choose $\delta_1 > 0$ so that for all $\omega_1, \omega_2 \in S^2$,
if $\|\omega_1 - \omega_2 \| < \delta_1$, then
\be
\takeanumber
\label{eq: p f a uniformly continuous}
\tag{\thetheorem}
\| p_{\hat{f},\hat{a}}(\omega_1) -
p_{\hat{f},\hat{a}}(\omega_2) \| < \frac{\epsilon}{2}.
\ee
We shall next choose $\delta_2$.  Now $\overline{\Theta}$
will mean
the $\delta_2$-thickening of $\partial B^{\hat{f},\hat{a}}$.
Choose $\delta_2$ so that
\begin{itemize}
\item
$\delta_2 < \epsilon/2$,
\item
$n_{\hat{f},\hat{a}}$ is defined
on $\overline{\Theta}$, and
\item
for all $x,y \in \overline{\Theta}$, if $\|x-y\| < \delta_2$,
then
\be
\| n_{\hat{f},\hat{a}}(x) - n_{\hat{f},\hat{a}}(y) \| < \frac{\delta_1}{2}.
\ee
\end{itemize}
\noindent
Choose $\delta$, with
\be
0 < \delta < \frac{\delta_1}{2}
\ee
so that for all $f,a \in N_\delta(\hat{f})\times
N_\delta(\hat{a})$,
\begin{itemize}
\item
$\partial B^{f,a} \subseteq \overline{\Theta} =
\overline{N_{\delta_2}}(\partial B^{\hat{f},\hat{a}})$,
\item
$n_{f,a}$ is defined throughout $\overline{\Theta}$, and
\item
for all $y \in \overline{\Theta}$,
$\|n_{f,a}(y)-n_{\hat{f},\hat{a}}(y) \|< \frac{\delta_1}{2}.$
\end{itemize}

\noindent
Given $\omega \in S^2$, let $x = p_{\hat{f},\hat{a}}(\omega)$
and $y = p_{f,a}(\omega)$.  Since
$y \in \partial B^{f,a}$, $y\in \overline{\Theta}$:
choose $z \in \partial B^{\hat{f},\hat{a}}$
so that $\|z-y\| \leq \delta_2 < \epsilon/2$.

\noindent
Then
\begin{gather*}
n_{f,a}(y) = \omega = n_{\hat{f},\hat{a}}(x)\\
\| n_{\hat{f},\hat{a}}(x) - n_{\hat{f},\hat{a}}(z) \|
=\\
\| n_{f,a}(y)- n_{\hat{f},\hat{a}}(z) \| \leq \\
\| n_{f,a}(y)- n_{\hat{f},\hat{a}}(y) \| +
\| n_{\hat{f},\hat{a}}(y) - n_{\hat{f},\hat{a}}(z) \| \leq 
2 \frac{\delta_1}{2} = \delta_1.
\end{gather*}
Let $\omega_1 = n_{\hat{f},\hat{a}}(x)$ and
$\omega_2 = n_{\hat{f},\hat{a}}(z)$, so
$x = p_{\hat{f},\hat{a}}(\omega_1)$
and
$z = p_{\hat{f},\hat{a}}(\omega_2)$.
Since $\| \omega_1 - \omega_2\| < \delta_1$,
\be
\| x-z \| < \frac{\epsilon}{2}
\ee
(Equation \ref{eq: p f a uniformly continuous}).
Hence, since $\|y-z\| < \epsilon/2$,
\be
\| x - y \| < \epsilon.
\ee
That is, for all $f,a$ in $U$,
\be
\| p_{f,a}- p_{\hat{f},\hat{a}} \|_{S^2} < \epsilon.\qed
\ee

\subsection{Pair descriptors}
Given a compact family $\cal G$ of convex bodies,
we are concerned with the family of Jordan curves
defined by the pre-seams arising from pairs $B_0,B_1$
of bodies.  We need to
show that these curves are $C^1$ manifolds.  But more is needed
since the space of Jordan curves is actually a space of
parametrisations.  For this reason, we define a
{\em pair descriptor} as a quintuple
\be
\psi = f_0, f_1, v_0, t, v_1
\ee
where $f_0, f_1 \in \cal G$, $v_0$ and $v_1$ are orthogonal
unit vectors, and $t$ is a nonnegative real number.
The idea is that this describes a pair $B_0, B_1$ of
bodies which may touch but not intersect properly,
where the shortest connecting line-segment lies along
the direction $v_0$, and $t$ is the shortest distance between
the two bodies.

\begin{definition}
\label{def: pair descriptors}
Given a compact family $\cal G$ of convex bodies,
$\Psi$ will denote the space of pair descriptors.
\be
\Psi =
{\cal G} \times {\cal G} \times S^2 \times [0,\infty) \times S^2 .
\ee
We make it a metric space by defining the distance between
two descriptors as the maximum of the five distances separating
their components.

A typical descriptor will be denoted
\be
\psi = f_0, f_1, v_0, t, v_1 .
\ee
From the descriptor, several entities can be derived:
the bodies $B_0, B_1$, the `centre' $a$ of $B_1$ (see below),
the pre-seam, its parametrisation $\phi \mapsto s(\phi)$, and so on.

We shall incorporate circumflexes
into the notation.  That is,
$\hat{\psi}$ will be a typical descriptor, and its components
will be
\be
\hat{f}_0, \hat{f}_1,
\hat{v}_0, \hat{t}, \hat{v}_1.
\ee
The derived entities will also carry circumflexes.
\end{definition}

\begin{figure}
\centerline{\includegraphics[height=2in]{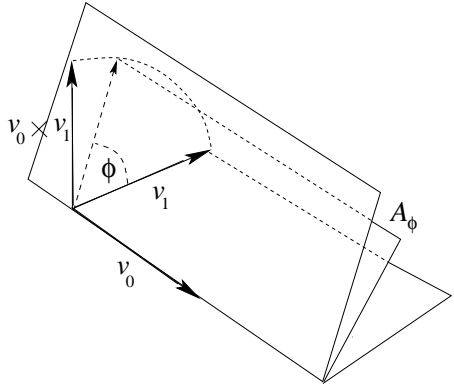}}
\caption{half-plane $A_\phi$ bounded by line through $v_0$.}
\label{fig: halfplane}
\end{figure}

The pre-seam is a subset of $S^2$;
the parametrisation $\phi\mapsto s(\phi)$ of the pre-seam
is fixed by choice of $v_0$ and $v_1$.  Let $v_2 = v_0 \times v_1$. Then
for any $0 \leq \phi \leq 2\pi$, the half-plane
\be
A_\phi = \{ \alpha v_0 + \beta ( v_1 \cos \phi  + v_2 \sin \phi ):
~ \alpha\in \IR, ~ \beta \geq 0 \}
\ee
contains a unique point on the pre-seam, and that point
is $s(\phi)$ (Figure \ref{fig: halfplane}.
The vector $v_1$ is almost arbitrary: it
defines a `starting direction.')

Clearly the properties of pre-seams are invariant under
simultaneous translation of the two bodies, so we can
assume that $B_0$ is `centred at $O$':
\be
B_0 =  \{ x: ~ f_0  (x ) \leq 1 \} .
\ee
The extreme point of $B_0$ in the direction $v_0$ comes
from the inverse normal map:
\be
p_{f_0,O}(v_0)
\ee
Suppose that $B_1=\{x:~ f_1(x-a) \leq 1\}$;
$B_1$ is `centred at $a$'.  Its extreme point
in the direction $-v_0$ is
\be
p_{f_1,O}(-v_0) + a
\ee
so
\be
p_{f_1,O}(-v_0) + a = p_{f_0,O}(v_0) + t v_0
\ee
whence
\be
\takeanumber
\label{eq: a, displacement of B1 in pair}
\tag{\thetheorem}
a = p_{f_0,O}(v_0) + t v_0 - p_{f_1,O}(-v_0) .
\ee

\begin{lemma}
\label{lem: a continuously dep}
The point $a$ just introduced depends continuously
on the descriptor $\psi$.
\end{lemma}

\noindent
{\bf Sketch proof.}
Let
\begin{gather*}
\hat{\psi} = \hat{f}_0, \hat{f}_1, \hat{v}_0, \hat{t}, \hat{v}_1
\\
\psi = f_0, f_1, v_0, t, v_1
\end{gather*}
be descriptors with associated
points $\hat{a}$ and $a$.  If $\hat{\psi}$ and
$\psi$ are sufficiently close together,
then
\begin{gather*}
\|p_{\hat{f}_0,O}(\hat{v}_0) - p_{f_0,O}(v_0) \|,\\
\|p_{\hat{f}_1,O}(-\hat{v}_0) - p_{f_1,O}(-v_0) \|,
\quad\text{and}\\
\|\hat{t}\hat{v}_0 -tv_0 \|
\end{gather*}
are all small, so $\|\hat{a}- a\|$ is small.\qed

\subsection{The pre-seam is a continuous Jordan curve on $S^2$}

In [\ref{hoy}, Lemma 5] it was proved 
that the pre-seam
is a Jordan curve --- homeomorphic to $S^1$ --- and a
parametrising map from $[0,2\pi]$ is given explicitly.
Without loss of generality
$v_0 = (1,0,0)$, and $v_1 = (0,1,0)$.\footnote{In [\ref{hoy}]
$(0,0,1)$ is the preferred direction for $v_0$.}

\subsection{The pre-seam is a $C^1$ manifold}

Recall that $p_{f,a}$ maps $\IR^3\backslash \{O\}$
onto $\partial B$, where $B = \{x:~ f(x-a) \leq 1\}$.
We shall omit $f,a$ and write $p$ alone, if no confusion arises.

\begin{lemma}
\label{lem: omega T p' omega = o}
For any $\omega\in \IR^3 \backslash \{ O \}$,
\be
\omega^T p'(\omega) = \vec{O}_{1\times 3} .
\ee
\end{lemma}

\noindent
{\bf Proof.}
(We use Fr\'{e}chet's definition
(\ref{par: frechet derivative})
of $p'(\omega)$.)
Let $x = p(\omega)$ so 
\be
\frac{\omega}{\|\omega\|} =
\frac{\nabla_f(x)}{\|\nabla_f(x)\|}
\ee
($\omega$ is not necessarily in $S^2$).
If $\omega + h \not= 0$, $f(p(\omega+h)) = f(p(\omega)) = 1$, so
\begin{gather*}
f(p(\omega+h))-f(p(\omega)) = 0\\
(\nabla_{f}(p(\omega)))^T p'(\omega) h \quad=\quad o( \| h \|) \\
(\nabla_{f}(p(\omega)))^T p'(\omega) = \vec{O}\\
\omega^T p'(\omega) = \vec{O}
\end{gather*}
\noindent since $\omega \propto \nabla_f(p(\omega))$.\qed

\hb

\begin{lemma}
\label{lem: pre-seam is a c1 manifold}
The pre-seam is a $C^1$ manifold.
\end{lemma}

{\bf Proof.}
We shall define a $C^1$ map $F: \IR^3\backslash \{O\} \to \IR^2$
and show that its derivative has rank 2 along the
pre-seam. It then follows from
the Implicit Function Theorem
[\ref{spivak}] that for any point $\omega$
on the pre-seam, projection onto one of the three
coordinate axes is a local $C^1$ diffeomorphism near
$\omega$.

The pre-seam (and a parametrisation) is specified by a pair descriptor
\begin{gather*}
f_0, f_1, v_0 , t, v_1 , \quad\text{where}\\
v_0 = (1,0,0)\quad\text{and}\quad v_1 = (0,1,0) ,
\end{gather*}
defining a pair $B_0,B_1$ of bodies whose closest points
are on the $x$-axis. Recall
(Equation \ref{eq: a, displacement of B1 in pair})
that 
\be
B_0 = \{x: f_0(x) \leq 1\}~\text{and}~
B_1 = \{x: f_1(x-a) \leq 1\}, \quad \text{where}~
a = p_{f_0,O}(v_0) + t v_0 - p_{f_1,O}(-v_0).
\ee
For simplicity, we write $p_0$ for $p_{f_0,O}$ and
$p_1$ for $p_{f_1,a}$.

Let $\omega$ be a point in $S^2$.  It is the
outward unit normal at exactly one point in $\partial B_0$
and one in $\partial B_1$, namely, $p_0(\omega)$ and
$p_1(\omega)$ respectively.
Let
\be
\takeanumber
\label{eq: q(omega)}
\tag{\thetheorem}
q(\omega) = p_1(\omega) - p_0 (\omega) .
\ee

$\omega$ is on the pre-seam if and only
if $p_0(\omega)$ is on the seam, or equivalently,
the (oriented) tangent plane $T$ to $\partial B_0$ at
$p_0(\omega)$ is also a supporting plane to $B_1$
at a point $y$.  But then $y = p_1(\omega)$.
So $p_1(\omega) \in T$, and $T$ is normal to $\omega$,
so $\omega$ is on the pre-seam if and only if
\be
\takeanumber
\label{eq: characterises pre-seam}
\tag{\thetheorem}
\omega^T q(\omega) = 0.
\ee
Therefore the pre-seam is the set of all $\omega \in S^2$
such that $\omega^T q(\omega) = 0$.

The map $F$ is
\be
F:~ \omega \mapsto ( \omega^T \omega, \omega^T q(\omega) ).
\ee

\noindent
By Equation \ref{eq: characterises pre-seam},
the pre-seam is $F^{-1}(1,0)$.


By a simple calculation, the derivatives
of $\omega^T\omega$ and $\omega^T q(\omega)$ are
\be
2\omega^T \quad\text{and}\quad
q(\omega) + \omega^T q'(\omega)
\ee respectively.
But
\be
\omega^T q'(\omega) = 
\omega^T p_1'(\omega) - \omega^T p_0'(\omega) = \vec{O}
\ee
(Lemma \ref{lem: omega T p' omega = o}).

Writing $F'$ as a $2\times 3$ matrix,
which is the correct format,
\be
F'(\omega)\quad=\quad
\left [ \begin{array}{c} 2\omega^T\\ q^T (\omega) \end{array}\right ]
\ee

All points in the pre-seam have unit length, so
near the pre-seam,
$\omega$ is nonzero, and  $q(\omega)$ is nonzero
since $B_0$ and $B_1$ can touch at one point
at most, and at that point the outward normals are opposite.
Also, if $\omega$ is on the
pre-seam then $\omega$ and $q(\omega)$ are orthogonal
(Equation \ref{eq: characterises pre-seam}).
Therefore $F'(\omega)$ has rank 2 near the pre-seam.
By the Implicit Function Theorem [\ref{spivak}],
the pre-seam is a $C^1$ manifold with local coordinate
systems provided by projection onto the coordinate
axes.\qed\hb

\noindent
For this application we can say more.

\begin{lemma}
\label{lem: y or z coordinate}
At any point $\omega$ in the
pre-seam, either the $y$- or the $z$-coordinate
is a local $C^1$ coordinate system.
\end{lemma}

{\bf Proof.}
Suppose $\omega$ is written with coordinates
$(x,y,z)$, and $q = (q_1,q_2,q_3)$.
The coordinates of $F'(\omega)$ are
\be
\left [ \begin{array}{ccc}
2x & 2y & 2z \\
q_1 & q_2 & q_3
\end{array}\right ] .
\ee

The $x,y,$ or $z$-coordinate gives a local coordinate system.
We would be obliged to use the
$x$-coordinate if the only choice of columns
with rank $2$ were the second and third.

But $B_0$ is left of $B_1$: $q_1 > 0$, so the first column
is nonzero and
it could be exchanged with one of the other two
to produce a linearly independent pair of columns, as
required.\qed\hb

\subsection{The pre-seam is a $C^1$ Jordan curve}

The pre-seam for a pair $B_0,B_1$ has a continuous parametrisation
$\{s(\phi): ~ 0 \leq \phi \leq 2\pi\}$.
We have a descriptor
\be
f_0, f_1, v_0, t, v_1
\ee
for the pair $B_0,B_1$.

\numpara
\label{par: implicit fun thm rel v0 v1 v0xv1}
Note: if we take coordinates $(\alpha, \beta, \gamma)$ relative
to the right-handed basis $v_0, v_1, v_0\times v_1$, the Implicit
Function Theorem can be interpreted relative to these three coordinate
axes, and from Lemma \ref{lem: y or z coordinate} we deduce:

\begin{corollary}
\label{cor: beta or gamma coordinate}
Near any point on the pre-seam, either
$\beta$ or $\gamma$
is, locally, a $C^1$ coordinate system for the pre-seam.\qed
\end{corollary}

\medskip

\noindent Again, there is little loss of generality in assuming
\be
v_0 = (1,0,0)\quad\text{and}\quad v_1 = (0,1,0):
\quad v_0 \times v_1 = (0,0,1).
\ee
Then there exists a unique $\theta$ such that
\be
s(\phi)  = (x,y,z) = (\sin \theta, \cos\theta \cos\phi,
\cos\theta\sin\phi) \in S^2,
\ee
where $-\pi/2 < \theta < \pi/2$ and $0 \leq \phi \leq 2\pi$.

More generally (relative to the basis $v_0, v_1, v_0\times v_1$),
\be
s(\phi) = \sin\theta ~~ v_0 + \cos\theta\cos\phi ~~ v_1
+ \cos\theta \sin\phi ~~ v_0 \times v_1 .
\ee

\begin{lemma}
\label{lem: x y not both zero}
Given $s(\phi) = (x,y,z)$ parametrised by $\theta$ and $\phi$,
$x \not= \mp 1$ (i.e.,
$\theta \not= \mp \pi/2$), and $y$ and $z$ are
not both zero.
\end{lemma}
{\bf Proof.}
Without loss of generality
$v_0 = (1,0,0)$ and $v_1 = (0,1,0)$.
The plane tangent to $B_0$ with outer normal $(-1,0,0)$
touches $B_0$ at its leftmost point and does not
touch $B_1$, so it is not a common tangent plane.
Similarly for the plane touching $B_1$ with outer
normal $(1,0,0)$.\qed

\medskip

Now fix $(x_0,y_0,z_0) = \omega_0 = s(\phi_0)$.
By Lemma \ref{lem: y or z coordinate}, projection onto the $y$- or $z$-axis
near $\omega_0$ is a local $C^1$ diffeomorphism and the inverse map
is a local coordinate system (for the pre-seam).  Without loss of
generality, the $y$-coordinate can be used as a
$C^1$ coordinate system.  That is, the map $y \mapsto (x,y,z)$
is a local coordinate system for the pre-seam.  We
can write $(x,y,z) = (g_1(y), g_2(y), g_3(y))$.

To express $\phi$ in terms of $y$:
\begin{gather*}
z = g_3(y)\\
\phi = \begin{cases}
\tan^{-1} ( z/y ) \quad\text{if}~ y \not= 0\\
\cot^{-1} (y/z) \quad\text{if} ~ y=0 .
\end{cases}
\end{gather*}

The functions $\tan^{-1}$ and $\cot^{-1}$ are understood
to have their domains and ranges adjusted
so that $\phi = \phi_0$ when $y = y_0$.

\begin{lemma}
\label{lem: case phi0 not O}
Given $\phi_0$, assume (w.l.o.g.)
that projection on the
$y$-axis is a local diffeomorphism of the
pre-seam near $s(\phi_0)$.

If $\phi_0 \not= 0$ then there is an interval
$(\phi_0 - \epsilon, \phi_0 + \epsilon)$ so
that the map
\be
\phi \mapsto y ( s ( \phi ) )
\ee
(that is, the $y$-coordinate of $s(\phi)$)
is a $C^1$ diffeomorphism onto an interval $(y_0-\delta, y_0+\eta)$.

Also, the map $\phi\mapsto s(\phi)$ is a local $C^1$ diffeomorphism
on the interval $(\phi_0 - \epsilon, \phi_0 + \epsilon)$.
\end{lemma}

\noindent
{\bf Proof.}
With little loss of generality, $y_0 \not= 0$.
There is an interval containing $y_0$ and a $C^1$-map
\be
y \mapsto \phi = \tan^{-1}(z/y) = \tan^{-1}(g_3(y)/y)
\ee
whose inverse is also $C^1$ (Inverse Function Theorem,
[\ref{spivak}]).
Since $\phi_0 \not= 0$, we can restrict the domain
of the inverse to $(\phi_0 - \epsilon, \phi_0 + \epsilon) \subseteq [0,2\pi]$
giving us a local $C^1$  diffeomorphism
\be
\phi \mapsto y
\ee
taking $\phi_0$ to $y_0$.  Now use $y$ as a coordinate system,
and compose maps, giving a local $C^1$ diffeomorphism
\be
\phi \mapsto y \mapsto (g_1(y), g_2(y), g_3(y) ) = s(\phi)
\ee
so $s$ is a local $C^1$ map near $\phi_0$.\qed

\begin{lemma}
\label{lem: case phi0 is O}
If $\phi_0 = 0$, and $s(\phi_0) = (x_0,y_0,z_0)$,
then there is an open subset of $[0,2\pi]$ of the form
\be
[0,\epsilon) \cup (2\pi - \epsilon, 2\pi]
\ee
and the restriction of $s$ to this set is (allowing
that $s(0) = s(2\pi)$) a local $C^1$ diffeomorphism.
(A messier version of the above lemma, proof omitted.)\qed
\end{lemma}

Summarising:

\begin{corollary}
\label{cor: C1 parametrisation}
The map $\phi \mapsto s(\phi)$ is a $C^1$ parametrisation of
the pre-seam: a $C^1$ Jordan curve.
\end{corollary}

\subsection{Bounded intersection number}

Recall that pre-seams are semi-algebraic of bounded degree
(Corollary \ref{cor: pre-seam semi-algebraic}).  It follows
that if two pre-seams are in general position, then
they intersect a bounded number of times:

\begin{lemma}
\label{lem: bounded intersection}
If $\cal G$ is a compact family of convex bodies, then
the associated family of
pre-seams has bounded intersection number
(Definition \ref{def: bounded intersection number}).\qed
\end{lemma}

\section{Displaced parameters}
\label{sect: displaced parameters}
A pair $\hat{B}_0, \hat{B}_1$ of bodies specified by
a pair descriptor 
\be
\takeanumber
\label{eq: hatted desc}
\tag{\thetheorem}
\hat{f}_0, \hat{f}_1, \hat{v}_0, \hat{t}, \hat{v}_1
\ee
the choice of $\hat{v}_0$ and $\hat{v}_1$ allow a unique parametrisation
$\phi \mapsto \hat{s}(\phi)$ of the $\hat{B_0}, \hat{B_1}$ pre-seam.
But in order to compare pre-seams derived from different
descriptors, we need to reconcile
their parametrisations.

As usual, one can assume that $\hat{v}_0 = (1,0,0)$ and
$\hat{v}_1 = (0,1,0)$.

\begin{figure}
\centerline{\includegraphics[height=1in]{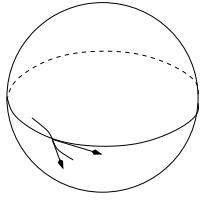}}
\caption{tangent to pre-seam is not aligned with $x$-axis.}
\label{fig: skew}
\end{figure}

\begin{lemma}
\label{lem: shallow angle}
Given bodies $\hat{B}_0, \hat{B}_1$
derived from a descriptor (\ref{eq: hatted desc}),
write $\hat{s}:~[0,2\pi] \to S^2$ for their pre-seam.
For any $\phi$, let $T_\phi$ be the tangent {\em line}
to $\hat{s}$ at $\hat{s}(\phi)$, let $P_\phi$ be the
{\em plane} through $T_\phi$ and $O$, and let
$\alpha(\phi)$ be the angle (between $0$ and $\pi/2$)
which this plane makes with the $x$-axis.
Let $\alpha=\inf_\phi \alpha(\phi)$.  Then $\alpha > 0$.
(See Figure \ref{fig: skew}).
\end{lemma}

{\bf Proof.} $P_\phi$ depends continuously on $\phi$,
so $\alpha(\phi)$ does also.
By compactness, it is minimised at some angle
$\phi_0$.  Let $\omega_0 = \hat{s}(\phi_0)$.

Suppose that $\alpha(\phi_0) = 0$.  Recall (Lemma
\ref{lem: pre-seam is a c1 manifold}) that there exists
a vector $q = [q_1,q_2,q_3]^T$, with $q_1 > 0$, such that
$T_{\phi_0}$ is orthogonal to $q$.
Also $q$ is orthogonal to $\omega_0$.
Thus $q$ is orthogonal to $P_{\phi_0}$ which contains
the $x$-axis. This means that $q$
is parallel to the $yz$-plane, whereas
$q_1 > 0$, a contradiction.\qed

\begin{lemma}
\label{lem: displaced}
With the same conditions as in Lemma \ref{lem: shallow angle},
let $\alpha$ be the minimum angle, as in the Lemma.
There exists an angle $\beta$ such that,
given orthogonal unit
vectors $v_0, v_1$, where the angle $v_0$ makes with the $x$-axis
is $<\beta$, the $\hat{B}_0, \hat{B}_1$ pre-seam can be uniquely
parametrised by angle $\rho$ around $v_0$:
$\rho \mapsto \tilde{s}(\rho)$, where $\tilde{s}(0)$ is in
the plane containing $v_0$ and $v_1$.
\end{lemma}

{\bf Proof.}
Write $S$ for the pre-seam $\hat{s}(\phi): ~ 0 \leq \phi \leq 2\pi$.
$S$ separates $S^2$ into two open sets $X\cup Y$
one containing $(-1,0,0)$ and the other $(1,0,0)$.

The
angle $\beta$ is chosen so that $\beta\leq\alpha$, and also
so that $-v_0 \in X$ and $v_0 \in Y$.

Let $A = \{t v_0: ~ t \in \IR\}$.

At no point point in $S$,
$\hat{s}(\phi)$, say, does the plane $P_\phi$, as defined in
the previous lemma, contain the line $A$.
Otherwise the angle between the $x$-axis and $P_\phi$,
would be bounded by that between the $x$-axis and $A$,
which is less than $\alpha$.  Equivalently:
the tangent line $T_\phi$ is not in the
plane containing $\hat{s}(\phi)$ and $A$.

For $0 \leq \rho \leq 2\pi$, let $A_\rho$ be the
half-plane, bounded by $A$, and at angle $\rho$ to the
half-plane containing $v_1$.

Suppose that for some $\rho$, $A_\rho$ intersects $S$
more than once.
By a variant of the Mean Value Theorem,
it could be rotated around $A$ into
a half-plane $A_{\rho'}$ tangent to S, which
is impossible. Hence every half-plane intersects
$S$ at most once.  Since every half-plane contains
$v_0 \in Y$ and $-v_0 \in X$, intersecting both $X$
and $Y$, it intersects their common boundary,
$S$ (Jordan Curve Theorem).\qed

\begin{lemma}
The above parametrisation $\rho \mapsto \tilde{s}(\rho)$
is continuous.
\end{lemma}

{\bf Proof.}  Given a sequence
a sequence $\rho_n \to \rho$, let $\omega_n = \tilde{s}(\rho_n)$
Let $x_n$ and $y_n$ be the points on $\partial \hat{B}_0$ and
$\partial \hat{B}_1$ respectively with outer unit normal $\omega_n$.
We can assume that all these sequences converge:
$\omega_n \to \omega \in S^2$, $x_n \to x \in \partial \hat{B}_0$,
and $y_n \to y \in \partial \hat{B}_1$.

It is enough to show that $\omega = \tilde{s}(\rho)$.

Let $T_n$ be the tangent plane to $\partial \hat{B}_0$ at $x_n$,
so $y_n \in T_n$.  Let $T$ be the tangent plane to $\partial \hat{B}_0$
at $x$.  Since $y_n \to y$, $y$ becomes arbitrarily
close to the planes $T_n$. For any $\epsilon > 0$,
let $T_\epsilon$ be a `thickening' of $T$: 
the slab consisting of all points
at distance $\leq\epsilon $ from $T$. For all sufficiently large
$n$, $y_n \in T_\epsilon$.  Therefore  $y\in T_\epsilon$, for
all $\epsilon > 0$, so $y \in T$, and $\omega$ is in the
pre-seam.  By a similar `thickening' argument, $\omega \in A_{\rho}$.
Therefore $\omega = \tilde{s}(\rho)$, as required.\qed

\section{The mapping to pre-seams is ($C^2\to\sup$)-continuous}
\label{sect: sup-contin}
Continuity in this sense means that if two pairs
$\hat{B}_0, \hat{B}_1$ and $B_0, B_1$ of bodies
are close under the $C^2$ metric, and $\hat{s}$ and
$s$ are the corresponding pre-seams, then
\be
\| s - \hat{s} \|_\infty
\ee
is small.

{\bf Proof strategy.}
We prove it in two stages.  First, we replace
$s$ by a `displaced' parametrisation $\tilde{s}$, which
uses the axes $\hat{v}_0, \hat{v}_1$, and show
\be
\| \tilde{s} - \hat{s} \|_\infty
\ee
is small; then we show that
\be
\| s - \tilde{s} \|_\infty
\ee
is small.

\begin{figure}
\centerline{\includegraphics[height=1.5in]{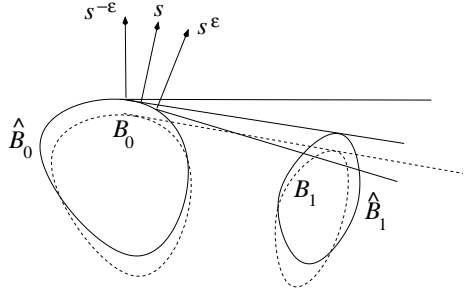}}
\caption{2-dimensional analogue.}
\label{fig: 2dim}
\end{figure}

The first stage uses a bracketing argument, which
is best explained by the two-dimensional problem,
in which the bodies are two-dimensional, the  sleeves
are pairs of line-segments, and the seams are pairs of
points. To show that the seams are within distance
$\epsilon$, it is enough to show that the (upper)
$B_0, B_1$ common tangent is sandwiched
between two lines $T^{\mp \epsilon}$ which are
close to the (upper) $\hat{B}_0, \hat{B}_1$ common tangent.
See Figure \ref{fig: 2dim}.

\hb

\noindent
We are given a pair descriptor
\be
\hat{\psi} = \hat{f}_0, \hat{f}_1, \hat{v}_0, \hat{t}, \hat{v}_1
\ee
(Definition \ref{def: pair descriptors}), with associated pre-seam
$\hat{s}$.  Given $\epsilon>0$, 
we want a neighbourhood $U$ of this descriptor
such that for all pre-seams $s$ derived from descriptors
in $U$,
\be
\|s-\hat{s}\|_\infty < \epsilon.
\ee

Given $-\pi/2 \leq \theta \leq \pi/2$ and
$0 \leq \phi \leq 2\pi$,
define
\be
\omega(\theta, \phi) =
( \sin\theta, \cos\theta \cos\phi, \cos\theta \sin \phi ) .
\ee
The map $(\theta, \phi) \mapsto \omega (\theta,\phi)$
is surjective.  It is not bijective because
$(\mp\pi/2,\phi) \mapsto (\mp 1,0,0)$ for all $\phi$,
but if $\theta \not= \pm \pi/2$ then $\phi$ is unique.

We know (Lemma \ref{lem: displaced})
that if $U$ is small enough then the pre-seams
can be parametrised continuously by angle $\phi$ around
$\hat{v}_0$.  Again we assume $\hat{v}_0 = (1,0,0)$ and
$\hat{v}_1 = (0,1,0)$.

\begin{definition}
Given 
\be
{\psi} = {f}_0, {f}_1, {v}_0, {t}, {v}_1,
\ee
yielding a pre-seam ${s}$, and $0 < \epsilon < 1$,
the pre-seam has positive distance from $(\mp 1,0,0)$, so
we may assume that for $0 \leq \phi \leq 2\pi$,
\be
\epsilon - 1 < (1,0,0)\cdot s(\phi) < 1 - \epsilon.
\ee
Choose a positive angle $\eta$ so that for all $\theta$ and $\phi$,
\be
\takeanumber
\tag{\thetheorem}
\label{eq: choice of eta}
\| (\sin(\theta+\eta), \cos(\theta+\eta)\cos\phi,
\cos(\theta+\eta)\sin \phi)
- (\sin\theta, \cos\theta\cos\phi, \cos\theta\sin \phi)
\| \leq \epsilon
\ee
It is sufficient that $\eta > 0$ and
$\sin \eta + \sin^2 (\eta/2) \leq \epsilon/2$,
and $\eta = \sqrt{2\epsilon/5}$ will do, if that
is less than $1$.
We define
\be
s^{\mp\epsilon} (\phi) = \omega(\theta \mp \eta, \phi)
\ee
where $s(\phi) = \omega(\theta, \phi)$.
\end{definition}
In other words, $s^{\mp \epsilon}$ is obtained by displacing
the pre-seam $s$ through angles $\pm \eta$ along lines
of constant $\phi$.
Note that by
choice of $\eta$, $s^{\mp \epsilon}$ define
$C^1$ Jordan curves, and $\| s^{\mp \epsilon} -s\|_\infty \leq \epsilon$.

We need to consider subsets $I_\delta$ of $\hat{B}_1$ which
have distance $> \delta$ from the boundary, or equivalently,
from the complement:

\begin{lemma}
\label{lem: strict interior}
Let $B = B^{f,a}$, and $\delta > 0$, such that
$\overline{N_\delta(a)} \subseteq B^\circ$, we define
$I_\delta$, which is an open subset of $B$, as
\begin{gather*}
I_\delta =
\{ x \in B: ~ d(x, \partial B) > \delta \}
=\\
\{ x \in B: ~ d(x, \IR^3 \backslash B) > \delta \}.
\end{gather*}
Then $I_\delta$ is nonempty and convex.\footnote{Its boundary
need not be differentiable.}

\noindent
Also, let
\be
O_\delta = \overline{N_\delta(B)} =  \{x: ~ d(x,B) \leq \delta\}
\ee
Then $O_\delta$, which is closed, is convex.
\end{lemma}

\noindent {\bf Proof.}
We need to show that for any $x,y \in I_\delta$,
the closed line-segment $xy$ is $ \subseteq I_\delta$.
Assume $x\not= y$, ignoring a trivial case.

Fix $z\in xy$ (the closed line-segment):
$z = (1-t)x + ty$ for some $t$ in $[0,1]$.
Now,
$z\in I_\delta$ if and only if
$\overline{N_\delta(z)} \subseteq B^\circ$.
Let $w$ be any point in $\overline{N_\delta(z)}$,
so $w = z + u$ where $\| u \| \leq \delta$.

The points $x+u$ and $y+u$ are both in $B^\circ$,
and so also is $(1-t)(x+u) + t(y+u) = z + u$.
Therefore $\overline{N_\delta(z)} \subseteq B^\circ$, as required.

For convexity of $O_\delta$, suppose
$x,y$ are at distance $\leq \delta$ from
$B$, and $z = (1-t)x + ty$ where $0 \leq t \leq 1$.
Choose $x',y' \in B$ at distance $\leq \delta$
from $x,y$, respectively. Let $z' = (1-t)x' + ty'$. Then
$z' \in B$, and
\be
\|z - z'\| = \| (1-t) (x-x') + t (y-y') \|
\leq (1-t) \| x-x' \| + t \| y-y' \| \leq \delta,
\ee
as required.\qed

Note that
\be
N_\delta(\partial B) = O_\delta \backslash I_\delta.
\ee
Sets like $I_\delta$ were considered in Lemma \ref{lem: shell thickening}.

\begin{figure}
\centerline{\includegraphics[height=2in]{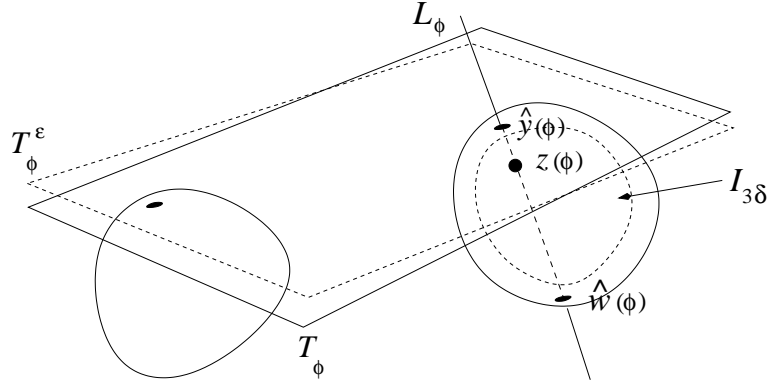}}
\caption{$z(\phi) \in L_\phi \cap I_{3\delta}$.}
\label{fig: 3delta}
\end{figure}

\begin{lemma}
\label{lem: w continuous}
Let $B$ be a convex body, $n$ its outer normal function.

Let $L(y)$, $y\in \partial B$, be the line through $y$
parallel to $n(y)$, and let $w(y)$ is the {\em other} point
where $L(y)$ meets $\partial B$.

Then the map 
\be
\partial B \to \partial B;\quad y \mapsto w(y)
\ee
is continuous.
\end{lemma}

\noindent{\bf Sketch proof.}
Fix $y \in \partial B$, and let $w = w(y)$. Since
$L(y)$ cuts $\partial B$ transversally at $y$,
$w \not= y$.  Let $d = \|w-y\|$. For any sufficiently small
$\epsilon$,
there exists a neighbourhood
$V$ of $y$ in $\partial B$,
depending on $n$ and $d$, such that for any $y' \in V$,
the line $L(y')$ intersects $N_\epsilon(w)$ within
$B^\circ$.  Then $\|w(y') - w\| < \epsilon$.\qed

\begin{lemma}
\label{lem: passes below}
Let $\hat{\psi}$ be a pair descriptor.
For $0 \leq \phi \leq 2\pi$, let $T_\phi$ (implicitly depending on
$\hat{\psi}$) be the common
tangent plane to $\hat{B}_0$ and $\hat{B}_1$ with outward normal
$\hat{s}(\phi)$. Let $\hat{y}(\phi)$ be the point
where $T_\phi$ touches $\hat{B}_1$, and let
\be
L_\phi
\ee
be the line through $\hat{y}(\phi)$ normal to $T_\phi$.

Let $T^{\epsilon}_\phi$
be the plane tangent to $\hat{B}_0$ with outward normal 
$\hat{s}^{\epsilon}(\phi)$.

For the purposes of Corollary
\ref{cor: nearby pre-seam, hatted axes} below,
we assume that  $\epsilon$ is reasonably small, so $\sec \eta \leq 2$.
($\eta$ was introduced in formula \ref{eq: choice of eta}).
In particular, $T^\epsilon_\phi$ intersects $L_\phi$ transversally.
Let $z(\phi)$ be the point of intersection:
\be
L_\phi \cap T^\epsilon_\phi = \{ z(\phi)\}.
\ee
See Figure \ref{fig: 3delta}.

Then: if $\epsilon$ is small enough, there exists
a $\delta>0$ so that for all $\phi \in [0,2\pi]$,
\be
z(\phi) \in I_{3\delta}.
\ee
\end{lemma}

\noindent {\bf Proof.}
Fix $\phi$.

Since $\hat{s}^\epsilon(\phi)$ is to the right of $\hat{s}(\phi)$,
the plane $T^{\epsilon}_\phi$ intersects $\hat{B}_1^\circ$, close
to $\hat{y}(\phi)$ if $\epsilon$ is small.  Since $\hat{B}_1$
meets $T_\phi$ from below (the side opposite
the outward normal at $\hat{y})$, the intersection is within a wedge between
the two planes, and this wedge contains $\hat{y}$.

Let $\hat{w}(\phi) = w(\hat{y})$, the other point where $L_\phi$
intersect $\partial \hat{B}_1$
(Lemma \ref{lem: w continuous}).

It follows
that $z(\phi)$ is in the open line-segment $L_\phi \cap \hat{B}_1^\circ$,
joining $\hat{y}(\phi)$ to $\hat{w}(\phi)$.

The function $\|z(\phi)-\hat{y}(\phi)\|$ is continuous,
so it has a positive lower bound $\ell_1$.
The function $\| z(\phi) - \hat{w}(\phi)\|$ is continuous,
so it has a positive lower bound $\ell_2$.
There is some freedom in choosing $\delta$;
\be
\delta = \frac{\min(\ell_1, \ell_2)}{3.1}
\ee
will do.\qed

\begin{lemma}
\label{lem: passes above}
This time let $T^{-\epsilon}_\phi$ be the plane tangent to $\hat{B}_0$
with outward normal $s^{-\epsilon}(\phi)$.  Then, if
$\epsilon$ is sufficiently small, there exists a $\delta > 0$
so that for all $\phi$, $d(T^{-\epsilon}_\phi, \hat{B}_1) > 3\delta$.\qed
\end{lemma}

\begin{corollary}
\label{cor: nearby pre-seam, hatted axes}
If $\epsilon$ is positive and sufficiently small, then there
exists a neighbourhood $U$ of $\hat{\psi}$ such that for
all $\psi \in U$,
\be
\| \hat{s} - \tilde{s} \|_{\infty} < \epsilon
\ee
(or, equivalently, $\leq \epsilon$),
where $\tilde{s}$ is a parametrisation of the $\psi$-pre-seam
relative to $\hat{v}_0$ and $\hat{v}_1$.
\end{corollary}

\noindent {\bf Proof.}
The functions $\hat{s}^{\mp \epsilon}$ are as in the previous
two lemmas (related to the pre-seam $\hat{s}$ derived from $\hat{\psi}$).
Apply the above two lemmas
getting a $\delta$ which satisfies both.

We assume that $\epsilon$
is sufficiently small so that the angle $\eta$ (see \ref{eq: choice of eta})
satisfies 
\be
\sec \eta \leq 2.
\ee

Choose the neighbourhood $U$ of $\hat{\psi}$ so that for all
$\psi \in U$, given $f_0, B_1,a$ are derived from $\psi$ and noting
$B_0, \hat{B}_0$ are centred at $O$,
\begin{itemize}
\item
$\|p_{f_0,O}-p_{\hat{f}_0,O}\|_\infty < \delta$
(Corollary \ref{cor: p sub f,a continuous}), and
\item
$ \partial B_1 \subseteq N_\delta ( \partial \hat{B}_1 )$
(Lemma \ref{lem: shell thickening}).
\end{itemize}
It is enough to show, for every $\psi\in U$ and angle $\phi$,
that
$\tilde{s}(\phi)$
is between $\hat{s}^{-\epsilon}(\phi)$ and
$\hat{s}^{\epsilon}(\phi)$.

Fix $\phi$. Let $\omega = \hat{s}(\phi)$ and
$\omega^{\mp \epsilon} = \hat{s}^{\mp \epsilon}(\phi)$.

\noindent
Let $\hat{y}(\phi) = p_{\hat{f}_1,\hat{a}}(\omega)$ and
let $L_\phi$ be the line through $\hat{y}(\phi)$
in the direction $\omega$, as in Lemma \ref{lem: passes below}.

Let $z_0(\phi)$ be the point where
$L_\phi$ intersects the plane $T^{\epsilon}_\phi$,
as in Lemma \ref{lem: passes below}.

\noindent
$T^\epsilon_\phi$ touches $\hat{B}_0$ at
$p_{\hat{f}_0, O} ( \omega^\epsilon ).$
Let $T$ be the parallel plane (with outer normal $\omega^\epsilon$)
which touches $B_0$ at
$p_{{f}_0, O} ( \omega^\epsilon )$.

In order to show that $s(\phi)$ is at or to the left of
$\hat{s}^{\epsilon}(\phi)$,
it is enough to show that the plane $T$ intersects
$B_1$.

Let $z_1(\phi)$ be the point where $T$ intersects $L_\phi$.
We want to show that $z_1(\phi) \in B_1$.

First we show that $\| z_1(\phi) - z_0(\phi) \| < 2\delta$.
For the plane $T_\phi^\epsilon$ touches $\hat{B}_0$ at
$p_{\hat{f}_0,O}(\omega^\epsilon)$, and $T$ touches $B_0$
at $p_{f_0,O}(\omega^\epsilon)$,
so these points are separated by a distance of $< \delta$.
Therefore
\be
d(T, T^{\epsilon}_\phi) < \delta.
\ee
If we take the points $z_0$ and $z_1$ and project them
orthogonally onto a line $L$ normal to these planes
(i.e., in the direction $\omega^\epsilon)$, we get two
points $z'_0, z'_1$ so $\| z'_1 - z'_0 \| < \delta$.
The lines $L$ and $L_\phi$ are at a relative angle
$\eta$, and projection reduces distance by a factor
$\cos \eta \geq 1/2$, so, as claimed,
\be
\| z_1 - z_0 \| < 2 \delta.
\ee

Recall
\be
\partial B_1 \subseteq
N_\delta (\partial \hat{B}_1 ) = O_\delta \backslash I_\delta.
\ee
There are three convex sets, nested, intersecting 
$L_\phi$ in nested intervals:

\begin{itemize}
\item
$O_\delta \cap L_\phi = (w_0,w_5)$, say, (an open line-segment), containing
\item
$B_1 \cap L_\phi = [w_1,w_4]$, containing
\item
$I_\delta \cap L_\phi = [w_2, w_3]$, say.
\end{itemize}

The important point is that $B_1 \cap L_\phi \supseteq
I_\delta \cap L_\phi$, and $z_1(\phi) \in [w_2,w_3]$.
Therefore
$z_1(\phi) \in B_1$, as required.

By  a similar calculation,
the plane with outer normal $\omega^{-\epsilon}$
touching $B_0$ has positive distance from $B_1$ and
therefore $\tilde{s}(\phi)$ is to the right of
$\hat{s}^{-\epsilon}(\phi)$.  Since these bounding points (unit
vectors in $S^2$) are
at distance $\leq \epsilon$ from $\hat{s}(\phi)$,
\be
\|\tilde{s}-\hat{s}\|_\infty \leq \epsilon.\qed
\ee

Continuing the discussion in 
Corollary \ref{cor: nearby pre-seam, hatted axes}:
there is a pre-seam $\hat{s}$ derived from a descriptor
$\hat{\psi}$.  We want to find a neighbourhood $U$ of $\hat{\psi}$
so that for every pre-seam $s$ derived from $U$,
\be
\| \hat{s} - s \|_\infty 
\ee
is small.  We know that if $\tilde{s}$ is $s$, but parametrised
relative to the $x$- and $y$-axes (as is $\hat{s}$), then 
\be
\| \hat{s} - \tilde{s} \|_\infty
\ee
is small. Given $\psi$ from which $s$ is derived:
\be
\psi = f_0, f_1, v_0, t, v_1,
\ee
write $M$  for the matrix
\be
M = \left [ \begin{array}{ccc} v_0 & v_1 & v_0 \times v_1
\end{array}\right ]
\ee
where the three vectors are stored as column vectors.
As usual $\hat{v}_0 = (1,0,0)$ and $\hat{v}_1 = (0,1,0)$,
and the corresponding matrix $\hat{M}$ is the identity.

Write $\hat{S}$ and $S$ for the images in $S^2$ of
$\hat{s}$ and $s$ respectively; and taking $\tilde{s}$ as above,
while $s\not= \tilde{s}$, they both have image $S$.

\begin{lemma}
\label{lem: M componentwise}
$M$ is a rotation matrix.
Suppose $v_0 = (1+ \alpha_0, \beta_0, \gamma_0)$ and
$v_1 = (\alpha_1, 1 + \beta_1, \gamma_1)$, where
the quantities $|\alpha_i|,|\beta_i|,|\gamma_i|$ are
bounded by $\delta < 0$.  Then
\be
| M - I |
\ee
(a $3\times 3$ matrix of absolute values)
is bounded componentwise by $3\delta$.

Therefore, if $\|v_0 - (1,0,0)\| < \delta$
and $\| v_1 - (0,1,0)\| < \delta$,
\be
| M - I |
\ee
is bounded componentwise by $3\delta$. It follows that
for any $\omega \in S^2$,
\be
\| M \omega -  \omega \| < 9\delta .
\ee

(Easy proof omitted.)\qed
\end{lemma}

\begin{figure}
\centerline{\includegraphics[height=1.5in]{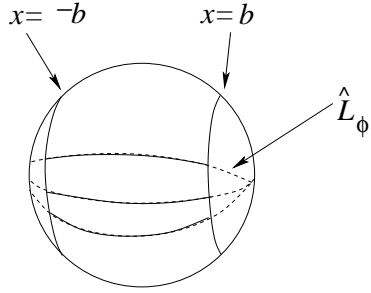}}
\caption{truncated sphere $S^2_b$.}
\label{fig: barrel}
\end{figure}

\begin{figure}
\centerline{\includegraphics[height=2in]{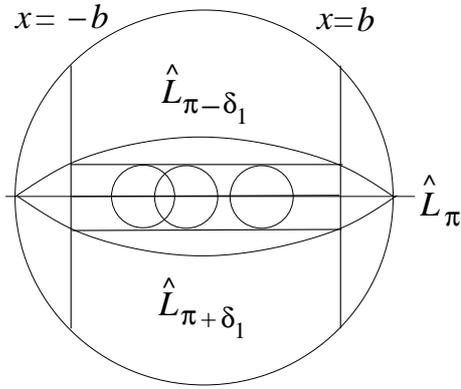}}
\caption{horizontal planes at height $\pm \sqrt{1-b^2} \sin \delta_1$.}
\label{fig: jamcake}
\end{figure}

\begin{lemma}
\label{lem: barrel staves}
Given $0 < b < 1$, the `truncated sphere' $S^2_b$
is
\be
\{ (x,y,z)\in S^2:~ -b \leq x \leq b\}
\ee
(see Figure \ref{fig: barrel}).
As usual, $\hat{v}_0 = (1,0,0)$ and $\hat{v}_1 = (0,1,0)$.
Write $\hat{L}_\phi$ for the great semicircle
\be
\hat{L}_\phi ~=~
\{ ( \sin \theta, \cos\theta \cos\phi, \cos\theta \sin\phi) :~
-\pi/2 \leq \theta \leq \pi/2\}.
\ee
Given $v_0, v_1$ as usual, write $M$ for the matrix
$[v_0~v_1~v_0\times v_1]$ as above, and write
$L_\phi$ for the great semicircle $M\hat{L}_\phi$.
Then:

For all (small positive) $\delta_1$ there exists
$\delta_2$ such that if $\| v_i - \hat{v}_i \| < \delta_2$
($i=0,1$) then (for any $\phi$, interpreting $\phi\mp \delta_1$
with wraparound at $2\pi$),
\be
L_\phi \cap S^2_b
\ee
is  between $\hat{L}_{\phi\mp \delta_1}$ in $S^2_b$.
\end{lemma}

\noindent {\bf Proof.}
Let
\be
R =
\{ (\sin \theta, \cos\theta\cos\phi', \cos\theta\sin\phi'):
~
-b \leq \sin\theta \leq b 
~\text{and}~
\phi-\delta_1 \leq \phi' \leq \phi+\delta_1
\}.
\ee

We need to show that if $\delta_2$ is sufficiently small and
$\|v_i - \hat{v}_i\| < \delta_2$ then
\be
M \hat{L}_\phi \cap S^2_b \subseteq R.
\ee

Without loss of generality $\phi=\pi$ and $\hat{L}_\phi$
is contained in the $xy$-plane (the `front': $y < 0$).

Take the two horizontal planes at heights $\pm \delta_3$, where
$\delta_3 = \sqrt{1-b^2} \sin \delta_1$.  Let $R'$ be
that part of $S^2_b$ contained between the two planes.
The $\delta_3$-neighbourhood in $\IR^3$
$N_{\delta_3}(\hat{L}_\phi)$
of $\hat{L}_\phi$ (see Figure \ref{fig: jamcake}) is
between these two planes, and its intersection with
$S^2_b$ is contained in $R'$.

$R$ and $R'$ have the same four corners.
Suppose $A$ is the upper horizontal plane.  The semicircle
$\hat{L}_{\phi-\delta_1}$ joins the upper two corners,
passing above $A$. Therefore the upper boundary of $R$
passes above $R'$.  Similarly with the lower boundary:
hence $R' \subseteq R$ and
\be
N_{\delta_3} (\hat{L}_\phi) \cap S^2_b ~\subseteq~ R.
\ee
Take $\delta_2 = \delta_3 / 9$.  Then for all $\omega$,
$\|M \omega - \omega \| \leq \delta_3$,
so
\be
L_\phi = M \hat{L}_\phi \subseteq N_{\delta_3} (\hat{L}_\phi)
\ee
so
\be
L_\phi \cap S^2_b \subseteq R
\ee
as required.\qed

\begin{corollary}
\label{cor: nearby pre-seam, correct axes}
Given a pre-seam $\hat{s}$ derived from a descriptor $\hat{\psi}$,
and $\epsilon>0$, there exists a neighbourhood $U$ of $\hat{\psi}$
such that for all pre-seams $s$ derived from descriptors in $U$,
\be
\| s-\hat{s} \|_\infty < \epsilon.
\ee
\end{corollary}

\noindent{\bf Proof.}
Write $\hat{S}$ for the image of $\hat{s}$, a Jordan
Curve.  Given $\epsilon > 0$, write
$\hat{S}^{\mp \epsilon/2}$ for the images of $\hat{s}^{\mp \epsilon/2}$.

Let $\hat{L}_\phi$ and $L_\phi$ have the same meanings as in Lemma
\ref{lem: barrel staves}.  In connection with that lemma,
we need to fix $b$ where $0 < b < 1$.

Assume $\epsilon$ is reasonably small, at most $1/10$, say.

\begin{itemize}
\item
Let $b = 1 -\epsilon/2$.
\item
Choose $\delta_1 > 0$ so that
for  all $\phi, \phi'$, if $|\phi - \phi' | < \delta_1,$
(allowing wraparound at $2\pi$),
then $\|\hat{s}(\phi) - \hat{s}(\phi') \| < \epsilon/2$.
\item
Choose $\delta_2 > 0$ so that for all $\phi$, if
$v_0, v_1$ are within distance $\delta_2$ of
$\hat{v}_0, \hat{v}_1$, respectively, and 
$L_\phi$ is the great semicircle at angle $\phi$
relative to the axes through $v_0$ and $v_1$, then
\be
L_\phi \cap S^2_b
\ee
is between $\hat{L}_{\mp \delta_1}$
(Lemma \ref{lem: barrel staves}).
\end{itemize}

\begin{figure}
\centerline{\includegraphics[height=1.5in]{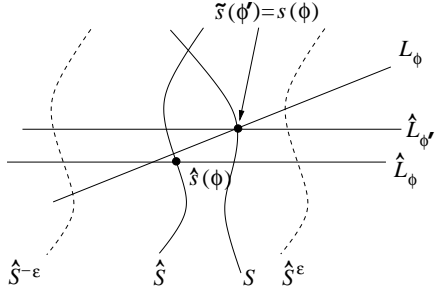}}
\caption{show that $s(\phi)$ is close to $\hat{s}(\phi)$.}
\label{fig: sshat}
\end{figure}

There exists a neighbourhood $U$ of $\hat{\psi}$ so
that for all $\psi$ in $U$, the image $S$ of
$s$ (the  pre-seam derived from $\psi$) is
between $\hat{S}^{\mp \epsilon}$
(Corollary \ref{cor: nearby pre-seam, hatted axes}).
We can also assume $U$ sufficiently small so that
for any such $\psi$, the derived vectors
$v_0, v_1$ are within distance $\delta_2$ of
$\hat{v}_0, \hat{v}_1$.
Given $0 \leq \phi \leq 2\pi$,
\begin{itemize}
\item
$\hat{L}_\phi \cap \hat{S} = \{ \hat{s} (\phi) \}$,
i.e., $\hat{s}(\phi)$ is the unique point common to
$\hat{L}_\phi$ and $\hat{S}$.
\item
$ \hat{L}_\phi \cap S = \{ \tilde{s} (\phi) \}$, and
\item
$ L_\phi \cap S = \{ s (\phi) \}$.
\end{itemize}
Now, $s(\phi) = \tilde{s}(\phi')$ for some unique
$\phi'$.   By Lemma \ref{lem: barrel staves}, since
$s(\phi) \in L_\phi\cap S^2_b$,
$\phi'$ is between $\phi-\delta_1$ and $\phi+\delta_1$,
so $\| \hat{s}(\phi') - \hat{s}(\phi) \| < \epsilon/2$.
Also, $\|\hat{s}(\phi') - \tilde{s} (\phi') \| < \epsilon/2$.
In other words, $\hat{s}(\phi') - s(\phi) \| < \epsilon/2$.
Therefore $\hat{s}(\phi) - s(\phi) \| < \epsilon$, as
required. See Figure \ref{fig: sshat}.\qed

\section{The mapping to pre-seams is ($C^2\to C^1$)-continuous}
\label{sect: c1-contin}
Having shown that the pre-seam map is
continuous from the $C^2$ metric to the sup metric,
it remains to prove:

\begin{lemma}
\label{lem: deriv to deriv continuous}
Given a descriptor $\hat{\psi}$, and $\epsilon > 0$,
there exists a neighbourhood $U$ of $\hat{\psi}$ such
that for every $\psi \in U$,
\be
\| \frac{ds}{d\phi} - \frac{d\hat{s}}{d\phi} \|_\infty < \epsilon
\ee
where $s$ and $\hat{s}$ are the pre-seams derived from
$\psi$ and $\hat{\psi}$ respectively.
\end{lemma}

{\bf Proof strategy.} We have invoked the Implicit Function
Theorem to show that the pre-seams are $C^1$ Jordan curves.
We strengthen this by showing that the derivative $ds/d\phi$
depends continuously on $s(\phi)$ and $q(s(\phi))$
the latter introduced in Equation \ref{eq: q(omega)}.

We need to review the calculations based on the
Implicit Function Theorem.  Suppose that a
descriptor $\psi$ derives the pre-seam.
Let $S$ be the pre-seam (a $C^1$ Jordan curve
in $S^2$) and $\phi\mapsto s(\phi)$ its parametrisation.

We write $v_0,v_1,v_2$ for the orthonormal basis,
$v_0, v_1$ given by $\psi$ and $v_2 = v_0 \times v_1$.
We write $(\alpha,\beta,\gamma)$ for coordinates with
respect to this basis; so if the basis is standard
then the coordinates coincide with $(x,y,z)$.

Recall that at any point $s(\phi_0)$ either
$\beta$ or $\gamma$ furnishes a local $C^1$
coordinate system. Furthermore, 
\be
\beta^2 + \gamma^2 > 0
\ee

Suppose that $\beta \not= 0$. Then
\begin{gather*}
\tan \phi = \frac{\gamma}{\beta}\\[3pt]
\frac{\beta \frac{d\gamma}{d\phi} - \gamma\frac{d\beta}{d\phi}}
{\beta^2}
=
\frac{d}{d\phi} \tan\phi =\sec^2\phi =\\[3pt]
\frac{\beta^2 + \gamma^2}{\beta^2}\\[3pt]
\end{gather*}
so
\be
\takeanumber
\label{eq: d beta and gamma}
\tag{\thetheorem}
\beta \frac{d\gamma}{d\phi} - \gamma\frac{d\beta}{d\phi}
= \beta^2 + \gamma^2.
\ee
The right-hand side is nonzero
at all points in $S$ (Lemma \ref{lem: x y not both zero}),
hence so is the left-hand side.  On the other hand,
if $\beta = 0$ so we use $\cot \phi = \beta/\gamma$,
we arrive at the same equation
(\ref{eq: d beta and gamma}).

Recall that the Implicit Function Theorem uses
\be
F ( \omega ) =
\left [ \begin{array}{c} \omega^T \omega \\ \omega^T q(\omega) \end{array}
\right ] =
\left [ \begin{array}{c} 1 \\  0 \end{array}\right ]
\ee
and $S = F^{-1} [1 ~ 0]^T$.  In this coordinate system
\be
F'(\omega) =
\left[ \begin{array}{ccc} 2 \alpha & 2 \beta & 2 \gamma \\
q_0 & q_1 & q_2
\end{array}\right ]
\ee
and  $ F'(\omega) ds/d\phi = O$.

We use $\omega = (\alpha,\beta,\gamma)$ and $w = (w_0,w_1,w_2)$ in
discussing the following matrix
\be
A \equiv A(\omega, w) =\left [ \begin{array}{ccc}
2\alpha & 2\beta & 2\gamma \\ w_0 & w_1 & w_2
\end{array}\right ]
\ee
where $\omega  = (\alpha,\beta,\gamma)$ and
$w = (w_0,w_1,w_2)$.  The condition $F'(\omega) ds/d\phi = O$ becomes
\be
\takeanumber
\label{eq: matrix A annulls}
\tag{\thetheorem}
A(s(\phi),q(s(\phi)))
\left [ \begin{array}{c}
\frac{d\alpha}{d\phi} \\[3pt]
\frac{d\beta}{d\phi} \\[3pt]
\frac{d\gamma}{d\phi}
\end{array}\right] = O.
\ee

The reason for indices $0,1,2$ is that the descriptors
$\psi$ furnish two vectors $v_0,v_1$, which we extend
to an orthornormal basis with $v_2 = v_0\times v_1$, and
the indexing is chosen to be consistent.

We label the $2\times 2$ minors of $A$ as
\be
g_0 =
\left | \begin{array}{cc} 2\beta & 2\gamma \\ q_1 & q_2 \end{array}\right |
,\quad
g_1 =
\left | \begin{array}{cc} 2\alpha & 2\gamma \\ q_0 & q_2 \end{array}\right |
,\quad\text{and}~
g_2 =
\left | \begin{array}{cc} 2\alpha & 2\beta \\ q_0 & q_1 \end{array}\right |
,
\ee
respectively.

Paraphrasing Lemma \ref{lem: y or z coordinate},
either (i) $g_1 \not= 0$ and $\beta$ can be used as
a local $C^1$ coordinate system, or (ii) $g_2 \not= 0$ and $\gamma$
can be used.

Suppose $g_1 \not= 0$. Then, from
Equation \ref{eq: matrix A annulls}, using Cramer's Rule,
\be
\frac{d\alpha}{d\phi} = \frac{- g_0 \frac{d\beta}{d\phi}}{g_1}
\quad\text{and}\quad
\frac{d\gamma}{d\phi} = \frac{- g_2 \frac{d\beta}{d\phi}}{g_1}
\ee
Substitute for $d\gamma/d\phi$ in Equation
\ref{eq: d beta and gamma}, and we have an equation
for $d\beta/d\phi$:
\be
\frac{d\beta}{d\phi} = \frac{-\beta^2 - \gamma^2}{\frac{g2}{g1}\beta + \gamma}.
\ee
Notice that the denominator on the right-hand side is nonzero,
since it is proportional to $\beta^2 + \gamma^2$.
We can substitute this to obtain similar expressions for
$d\alpha/d\phi$ and $d\gamma/d\phi$, and hence
\be
\frac{ds}{d\phi} = G_1 (v_0,v_1, s(\phi), q(s(\phi))
\ee
where $G_1(v_0, v_1, \omega, w)$
is a mildly complicated expression in these parameters.

The relation between $\alpha,\beta,\gamma$ and $\omega$
is
\be
\alpha = \omega^T v_0,\quad
\beta = \omega^T v_1,\quad\text{and}\quad
\gamma = \omega^T \, v_0\times v_1.
\ee
Also,
\be
\frac{d\alpha}{d\phi} = v_0^T G_1(v_0, v_1, s(\phi), q(s(\phi))),
\ee
with similar expressions for $d\beta/d\phi$ and $d\gamma/d\phi$.

$G_1$ is defined on the set 
\be
\{(v_0,v_1,\omega,w) \in S^2\times S^2
\times S^2 \times \IR^3:
g_1 (v_0, v_1, \omega,w) \not= 0\}.
\ee

When $g_2 \not= 0$ we get a similar expression
\be
\frac{ds}{d\phi} = G_2 (v_0, v_1, s(\phi), q(s(\phi)) .
\ee

In other words, since the pre-seam depends
continuously on $\psi$, and we have $ds/d\phi$
defined in terms of continuous functions $G_1, G_2$,
Lemma \ref{lem: deriv to deriv continuous} can be reduced to the
following:

\begin{lemma}
\label{lem: deriv to deriv continuous, version 2}
Given a descriptor $\hat{\psi}$ from which
a pre-seam $\hat{S}$ and its parametrisation
$\phi\mapsto \hat{s}(\phi)$ is derived, there exist open
sets $V_1,\ldots V_n$ (in $\IR^3$) covering
$\hat{S}$, and for each $V_i$, 
a neighbourhood $U_i$ of $\hat{\psi}$, and
a choice $k_i = 1$ or $k_i = 2$ such
that
for all $\omega,\hat{\omega} \in V_i$ and $\psi$ in $U_i$,
both terms $G_{k_i}(\ldots)$ given below are well-defined, and
\be
\|
G_{k_i} (v_0, v_1, \omega, q(\omega) )
-
G_{k_i} (\hat{v}_0, \hat{v}_1, \hat{\omega}, \hat{q}(\hat{\omega}) )
\| < \epsilon.
\ee
\end{lemma}

\noindent {\bf Proof.}
Fix $\phi_0$. Let $\hat{s}(\phi_0) = \hat{\omega}_0$.
Let $\hat{w}_0 = \hat{q}(\hat{\omega}_0)$.
Without loss  of generality
\be
G_1 (\hat{v}_0,\hat{v}_1, \hat{\omega}_0, \hat{w}_0 )
\ee
is  well-defined. Take neighbourhoods $Z_0$ of
$(\hat{v}_0, \hat{v}_1)$, $V_0$ of
$\hat{\omega}_0$ (in $S^2$) and $W_0$ of $\hat{w}_0$ in $\IR^3$
so that $G_1(v_0, v_1, \omega, w)$ is
well-defined and continuous throughout
$Z_0 \times V_0 \times W_0$, and
\be
\takeanumber
\label{eq: small variation G1 on V0}
\tag{\thetheorem}
\| G_1 ( v_0, v_1, \omega, w )
-
G_1 ( \hat{v}_0, \hat{v}_1, \hat{\omega}_0, \hat{w}_0 )
\| < \frac{\epsilon}{2}.
\ee

Shrink $V_0$ if necessary to a smaller neighbourhood
of $\hat{\omega}_0$, so that $\overline{V_0}$ is compact and
\be
\hat{q} ( \overline{V_0} ) \subseteq W_0.
\ee

For every $\omega \in \overline{V_0}$ there exists $\delta > 0$
so that $N_{2\delta}(\hat{q}(\omega)) \subseteq W_0$.
Choose $\omega_1,\ldots , \omega_m$ and
positive numbers $\delta_1,\ldots, \delta_m$, so that the finite
union
\be
\bigcup
S^2 \cap N_{\delta_j} ( \hat{q}(\omega_j))
\ee
covers the compact set $\hat{q}\overline{V_0}$.
 Let
$\eta_0 $ be the minimum of $\delta_1, \ldots, \delta_m$.

For any $\omega \in V_0$,
$\hat{q}(\omega) \in N_{\delta_j}(\omega_j)$
for some $j$, and 
\be
N_{\eta_0} (\hat{q}(\omega)) \subseteq W_0 .
\ee
For some neighbourhood $U_0$ of $\hat{\psi}$,
\be
\sup_{\omega\in S^2} \|q(\omega) - \hat{q}(\omega)\| < \eta_0,
\ee
(Equation \ref{eq: q(omega)} and Corollary \ref{cor: p sub f,a continuous})
and also
\be
(v_0, v_1) \in Z_0 .
\ee
Given $\psi$ in $U_0$,
fix $\omega,\hat{\omega}$ in $V_0$.  Write
$w$ and $\hat{w}$ for $q(\omega)$ (derived from $\psi$)
and $\hat{q}(\hat{\omega})$.
Since $\omega\in V_0$, $N_{\eta_0} (\hat{q} (\omega)) \subseteq W_0$.
Since $\psi\in U_0$, $\| q(\omega) - \hat{q}(\omega) \| <
\eta_0$. Therefore, 
\be
q(\omega) = w \in W_0
\ee
Also, of course, $\hat{q}(\hat{\omega}) \in W_0$.
Therefore 
\be
(v_0, v_1, \omega, q(\omega))
\in Z_0 \times V_0 \times W_0,
\ee
so
\be
\| G_1 (v_0, v_1, \omega, q(\omega) )
- G_1 (\hat{v}_0, \hat{v}_1, \hat{\omega}_0, \hat{w}_0 )
\| < \frac{\epsilon}{2},
\ee
as required.
Also,
\be
\| G_1 (v_0, v_1, \hat{\omega}, \hat{q}(\hat{\omega}) )
- G_1 (\hat{v}_0, \hat{v}_1, \hat{\omega}_0, \hat{w}_0 )
\| < \frac{\epsilon}{2}.
\ee
Therefore
\be
\| G_1 (v_0, v_1, \omega, q(\omega) 
-
G_1 (\hat{v}_0, \hat{v}_1, \hat{\omega}, \hat{q}(\hat{\omega}))
\| < \epsilon.
\ee

The result has been established locally at
$\hat{s}(\phi_0)$. By routine compactness arguments
we get a suitable open cover $V_1,\ldots, V_n$
and open neighbourhoods $U_1,\ldots, U_n$ of
$\hat{\psi}$.  The only difference is that
the number $\eta_0$ be replaced by the
minimum of the numbers $\eta_i$, and the set
$U_0$ be replaced by the neighbourhood
$U = U_1 \cap \ldots \cap U_n$.\qed

Combining Corollary
\ref{cor: nearby pre-seam, correct axes},
with Lemma \ref{lem: deriv to deriv continuous}
we conclude

\begin{theorem}
\label{thm: c2 to c1 continuous}
Given a compact family $\cal G$ of convex bodies,
the pre-seam map on the space of descriptors,
\be
\psi \mapsto s(\psi)
\ee
is continuous from the product metric on the space
of descriptors to the $C^1$ metric on $C^1$ Jordan
curves in $S^2$.\qed
\end{theorem}


\section{Pre-seams form a compact family}
\label{sect: pre-seam compact family}
In this section, $\cal G$ is a compact family of convex
bodies, $\Psi$ is the space of pair descriptors
from $\cal G$, and and $\cal F$ is the family of
pre-seams from $\cal G$ --- i.e., pre-seams derived from
descriptors in $\Psi$.

We shall prove that $\cal F$ is compact under the $C^1$
metric. Since $\cal F$ is a compact space, it  is enough
to prove that $\cal F$ is sequentially compact, a property
already mentioned in Section \ref{sect: metric}.

\begin{definition}
\label{def: seq compact}
A metric space $X$  is {\em sequentially compact}
if every infinite sequence $x_n$ of points in
$X$ has a convergent subsequence:
i.e., there exists an infinite subsequence $x_{n_i}$
and a point $x \in X$ such that
\be
\lim_{i\to \infty} x_{n_i} = x.
\ee
For metric spaces, compactness and sequential compactness
are equivalent.
\end{definition}
Thus we need to prove that, given an infinite sequence
$s_n$ of pre-seams, there exists a 
subsequence $s_{n_i}$ converging to a pre-seam $s$.

First choose a sequence $\psi_n$ of pair descriptors
such that for each $n$, $s_n$ is derived from $\psi_n$.
Write

\be
\psi_n \, = \, f_{0,n} \, f_{1,n} \, v_{0,n} \, t_n \, v_{1,n}
\ee

{\bf Proof strategy.}  If the parameters $t_n$ are bounded,
then we invoke the continuity of the pre-seam map.  If
the parameters $t_n$ are unbounded, then we can choose
a subsequence where the pre-seams converge to a great
circle on $S^2$, and the latter is also a pre-seam.
We use bracketing arguments: given any convex body
(since $\cal G$ is compact), one can always inscribe a
sphere of radius $m > 0$ and circumscribe a sphere of
radius $M < \infty$.

\begin{lemma}
\label{lem: tn bounded}
If the sequence $t_n$ is bounded, then
the sequence $s_n$ contains a subsequence converging
to a pre-seam $\hat{s}$ under the $C^1$ metric.
\end{lemma}

{\bf Proof.} Given that
$t_n \in [0,u]$ for
some $u$, then all $\psi_n$ belong to a compact subspace
of $\Psi$,  namely,
\be
{\cal G} \times {\cal G} \times S^2 \times [0,u] \times S^2,
\ee
so it admits a subsequence $\Psi_{n_i}$ converging to
a descriptor $\hat{\psi}$ in $\Psi$, and
since the pre-seam map is continuous, the derived
pre-seams $s_{n_i}$ converge to the pre-seam $\hat{s}$
derived from $\hat{\psi}$.\qed

We need only consider the case where
the $t_n$ are unbounded.  By passing to a subsequence
if necessary, we can assume $t_n \to \infty$, or indeed that
\be
t_n \geq n
\ee
for all $n$.

\begin{lemma}
\label{lem: inscribed, circum}
There exist positive real numbers $m,M$ such that
for all bodies $B$, of the form $\{x: f(x) \leq 1\}$,
where $f\in \cal G$,
\be
\overline{N_m(O)} \subseteq B \subseteq \overline{N_M(O)}.
\ee
\end{lemma}

{\bf Proof.} Since $f(x) = 2$ for $\| x \| \geq 1.5$,
we can take $M=1.5$.  To prove existence of $m$, we
recall that $O \in B^\circ$ for all such $B$.  If
such an $m$ does not exist, then $O$ can be
arbitrarily close to $\partial B$, and
there exists a sequence
$B_n$ of bodies derived from $f_n \in \cal G$, and
a body $B$ derived from $f$, where $f_n \to f$ in $\cal G$,
and a sequence $x_n \in \partial B_n$, where
$\| x_n \| \to 0$. Therefore $x_n \to O$, and since $f$ is continuous,
$f(x_n) \to 0$.  For sufficiently large $n$, $f(x_n) < 1/3$
and $|f_n(x_n) - f(x_n)| <  1/3$, and $f_n (x_n) = 1$,
which is impossible.\qed

There is something in common between proving
continuity of the pre-seam map around a 
descriptor $\hat{\psi}$, and what we need
to prove about limits. To underline the
connection, we use circumflexes in some places.

Given orthonormal  vectors $\hat{v}_0, \hat{v}_1$, let
\be
\hat{S} = \{ \beta \hat{v}_1 + \gamma  \,
\hat{v}_0\times \hat{v}_1:~ \beta^2 + \gamma^2 = 1 \}.
\ee
$S$ is the great circle normal to $\hat{v}_0$ in $S^2$.

\begin{lemma}
\label{lem: always a pre-seam}
The set $\hat{S}$ is a pre-seam, and its
parametrisation is
\be
\hat{s} (\phi) = \cos\phi \, v_0 +
\sin\phi \,  v_0\times v_1.
\ee
\end{lemma}

{\bf Proof.}
Let
\be
\hat{\psi} = f_0, f_0, \hat{v}_0, 100, \hat{v}_1.
\ee
This describes two widely-separated identical copies
of the same convex body.  The common tangent planes
are parallel to the direction $v_0$, so the
outer normals constitute the set $\hat{S}$,
and the parametrisation is $\phi\mapsto \hat{s}(\phi)$.\qed

Given a small positive $\epsilon$: $\epsilon \leq 1/2$, say,
let
\be
S^{\mp \epsilon} = 
\{ (\alpha \hat{v}_0 + \beta \hat{v}_1 + \gamma \, \hat{v}_0\times \hat{v}_1) \in S^2:~
\alpha = \mp \epsilon\}.
\ee

\begin{lemma}
\label{lem: hat S is s pre-seam}
For every $\epsilon > 0$ there exists an $\ell_\epsilon$ such
that for every descriptor in which the bodies are
sufficiently widely separated, and the basis
vectors are $\hat{v}_0$ and $\hat{v}_1$, i.e.,
\be
\psi= f_0, f_1, \hat{v}_0, t, \hat{v}_1
\quad\text{with}~ t \geq \ell_\epsilon,
\ee
the derived seam $S$ is between $\hat{S}^{\pm \epsilon}$.
\end{lemma}

{\bf Sketch proof.} Coordinates $\alpha, \beta, \gamma$
are with respect to the  basis $v_0, v_1, \, v_0\times v_1$.
Let $B_0, B_1$ be the bodies as  usual, with
$B_1$ centred at $a$ (which  depends on $t$). The pre-seam
is sandwiched between the
pre-seam for $\overline{N_m(O)},\overline{N_M(a)}$
and $\overline{N_M(O)}, \overline{N_M(a)}$, which
are two circles normal to $v_0$; if $t$,
and hence $\| a \|$, is sufficiently
large then these circles are between $\hat{S}^{\mp \epsilon}$.\qed

This result has much in common with Corollary
\ref{cor: nearby pre-seam, hatted axes}, which
was developed into Corollary \ref{cor: nearby pre-seam, correct axes}.
The following corollary is also  related.

\begin{corollary}
\label{cor: limit under sup norm}
Given
\be
\psi_n = f_{0,n} f_{1,n} v_{0,n} t_n v_{1,n}
\ee
where $v_{0,n}\to \hat{v}_0$, $v_{1,n} \to \hat{v}_1$,
and $t_n\to \infty$, the derived pre-seams $s_n$ converge
to $\hat{s}$:
\be
\lim_n \| s_n - \hat{s}\|_\infty = 0.
\ee
\end{corollary}
\noindent {\bf Proof omitted.}\qed

We turn to the $C^1$ norm.
Again, the pre-seam $s_n$ derived from $\psi_n$
has the form $F_n^{-1}(1,0)$, and the derivative
$F_n'(\omega)$ is 
\be
\left [ \begin{array}{ccc}
2\alpha_n & 2 \beta_n & 2 \gamma_n \\
q_{0,n} & q_{1,n} & q_{2,n}
\end{array} \right ]
\ee
and the important relation is $F_n'(s_n(\phi)) ds_n/d\phi = O$.
The bottom row grows with $n$, but if we {\em normalise}
it, we get a matrix in which all entries are bounded:
\be
\left [ \begin{array}{ccc}
2\alpha_n & 2 \beta_n & 2 \gamma_n \\
\overline{q}_{0,n} & \overline{q}_{1,n} & \overline{q}_{2,n}
\end{array} \right ]
\ee
where
\be
\overline{q}_{0,n} , \overline{q}_{1,n} , \overline{q}_{2,n})
= \overline{q_n} = q_n/\|q_n\|.
\ee

Recall that when discussing $C^1$ continuity of the pre-seam map,
we arrived at an equation
\be
A(s(\phi),q(s(\phi)) \frac{ds}{d\phi} = O
\ee
Scaling the bottom row does not change the nullspace, so
equivalently
\be
A(s(\phi),\overline{q}(s(\phi))) \frac{ds}{d\phi} = O
\ee
and we can base our calculations on this equation.
The bottom rows $\overline{q}_n $ converge
to $(1,0,0)$. Fix $\phi$.  We can write
$s_n(\phi)$ as $(\alpha_n, \beta_n , \gamma_n)$,
the subscripts $n$ indicating the basis in which
these coordinates are computed. Since
$v_{0,n}\to (1,0,0)$ and $v_{1,n}\to (0,1,0)$,
$(\alpha_n, \beta_n, \gamma_n) \to (x,y,z)$.

One may note in passing that (for any pre-seam $s$)
\be
\frac{ds}{d\phi} \propto s(\phi)\times \overline{q}.
\ee
It follows that $\lim s_n(\phi) \propto \hat{s}(\phi)$
but equality needs to be settled.

Recall that $g_0, g_1, g_2$ are the three minors
of the matrix $A(\omega,q)$, and we can let them
denote instead the minors of the matrix
$A(\omega, \overline{q})$.

Since $\| q_n \| \to \infty$, with unbounded separation in the $x$-direction
but not in the other directions, 
\be
\overline{q}_n \to (1,0,0).
\ee

The matrix $A$ for $\hat{s}$ is
\be
\left [ \begin{array}{ccc}
x & y & z \\ 1 & 0 & 0
\end{array}\right ]
\ee
and for this matrix, $g_1 = -z$, nonzero except where $z=0$,
$\phi = \mp \pi/2$..
If $s_n$ is sufficiently close to $\hat{s}$ then
$g_1$ is also nonzero, and we can calculate
(with $s_n(\phi) = (\alpha_n, \beta_n, \gamma_n)$ in
the appropriate coordinate system),
\be
\frac{d \beta_n}{d\phi} =
\frac{-\beta_n^2 - \gamma_n^2}
{
\frac{
\left | \begin{array}{cc} 2\alpha_n & 2\beta_n \\
\overline{q}_{0,n} & \overline{q}_{1,n} \end{array}
\right | }
{
\left | \begin{array}{cc} 2\alpha_n & 2\gamma_n \\
\overline{q}_{0,n} & \overline{q}_{2,n} \end{array}
\right | }
\beta_n+ \gamma_n
} .
\ee

If we take limits, so $\beta_n \to y$ and so on, we get
\be
\frac{-y^2 - z^2}{
\frac{-2y}{-2z} y + z } = -z
\ee
which is $d\hat{s}/d\phi$, as required.

A similar analysis applies when $z=0$ so $y= \mp 1$.
In conclusion,

\begin{lemma}
\label{lem: tn unbounded}
Suppose 
\be
\psi_n = f_{0,n} f_{1,n} v_{0,n} t_n v_{1,n}
\ee
where $t_n \to \infty$ and the other components converge,
to $\hat{f}_0, \hat{f}_1, \hat{v}_0, \hat{v}_1$ respectively. Let
\be
\hat{\psi} = \hat{f}_0, \hat{f}_0, \hat{v}_0, 100, \hat{v}_1
\ee
Then $s_n \to\hat{s}$ in the $C^1$ metric.\qed
\end{lemma}

Combining Lemmas \ref{lem: tn bounded} and \ref{lem: tn unbounded},
we have

\begin{corollary}
\label{cor: pre-seams compact family}
Given a compact family $\cal G$ of convex bodies,
with associated space $\Psi$ of pair descriptors,
every infinite sequence $s_n$ of pre-seams contains
a subsequence converging to a pre-seam $\hat{s}$.

Hence the family $\cal F$ of pre-seams is a compact
family of Jordan curves.\qed
\end{corollary}

\section{The main theorem}
\label{sect: main theorem}
\begin{theorem}
\label{thm: main theorem}
Let $\cal G$ be a compact family of convex bodies
in $\IR^3$ (with semialgebraic boundaries).  Then
for any set $S$ of $n$ disjoint bodies which are 
translations of bodies derived from $\cal G$,
the convex hull $H(S)$ has $O(n^2 \lambda_s (dn))$ features,
where $s$ and $d$ are constants depending on $\cal G$.
\end{theorem}

{\bf Proof.} The feature complexity of $H(S)$ is proportional
to the feature complexity of unions of hidden regions; for
each body $B$ in $S$, this is $O(\lambda_s (dn))$
from Corollary \ref{cor: bigcup Di has feature complexity}.\qed

\section{References}
\label{sect: references}
\begin{enumerate}

\item
\label{acv}
Helmut Alt, Otfried Cheung, and Antoine Vigneron (2005).
The Voronoi diagrams of curved objects.
{\em Discrete and Computational Geometry, \bf 34}, 439--453.

%


\item
\label{bcddy}
Jean-Daniel Boissonnat, Andr\'e C\'er\'ezo, Olivier
Devillers, Jacqueline Duquesne, and Mariette Yvinec (1996).
An algorithm for constructing the convex hull of a set of
spheres in dimension $d$.
{\em Computational Geometry: Theory and Applications
\bf 6:2}, 123--130.
\item
\label{dck}
Mark de Berg, Otfried Cheong, and Marc van Kreveld (2008).
{\em Computational geometry: algorithms and applications.}
Springer, 3rd edition.
\item
\label{ek}
Alon Efrat and Matthew J.\ Katz (1999).
On the union of $\kappa$-curved objects.
{\em Computational Geometry \bf 14}, 241--254.

\item
\label{bpr}
Saugata Basu, Richard Pollack, and Marie-Fran\c{c}oise Roy (2003).
{\em Algorithms in Real Algebraic Geometry}, Springer Series on
Algorithms and Computation in Mathematics.

\item
\label{gp}
Victor Guillemin and Alan Pollack (1974).
{\em Differential Topology}.  Prentice-Hall.

\item
\label{hoy}
Paul Harrington, Colm \'O D\'unlaing, and Chee-Keng
Yap (2007).
Optimal Voronoi diagram construction with $n$ convex sites in three dimensions,
{\em International Journal of Computational Geometry and Applications,
\bf 17:6}, 555--593.


\item
\label{hi}
C.-K.\ Hung and D.\ Ierardi (1995). Constructing convex hulls
of quadratic surface patches. {\em Proc 7th Canadian Conf. on
Computational Geometry}, 255--260.


\item
\label{ss}
Jacob T.\ Schwartz and Micha Sharir (1990). On the
2-dimensional Davenport-Schinzel problem.
{\em J. Symbolic Computation, \bf 10}, 371--393.
\item
\label{as}
Micha Sharir and Pankaj Agarwal (1995).
{\em Davenport-Schinzel sequences and their
geometric applications.}  Cambridge University Press.
\item
\label{spivak}
Michael Spivak (1998). {\em Calculus on manifolds.}
Addison-Wesley.

\item
\label{spivak2}
Michael Spivak (1999).  {\em A comprehensive introduction to
differential geometry, \bf I}. Publish or Perish.

\item
\label{stillwell}
John Stillwell (1980). {\em Classical topology and combinatorial group
theory.} Springer Graduate Texts in Mathematics {\bf 72.}

%
\item
\label{wolpert}
Nicola Wolpert (2002).  An exact and efficient approach for computing
a cell in an arrangement of quadrics. Doctoral dissertation,
University of the Saarland, Saarbr\"ucken.
\end{enumerate}

\end{document}